\documentclass[traditabstract,longauth]{aa}

\usepackage{graphicx}
\usepackage{txfonts}
\usepackage{natbib}
\usepackage[colorlinks=true,citecolor=blue]{hyperref}
\usepackage{xspace}
\usepackage{subfig}
\usepackage{xcolor}

\usepackage[T1]{fontenc}

\bibpunct{(}{)}{;}{a}{}{,}

\newcommand{\MJup}{\ensuremath{M_{\mathrm{Jup}}}\xspace}
\newcommand{\RJup}{\ensuremath{R_{\mathrm{Jup}}}\xspace}
\newcommand{\MSun}{\ensuremath{M_{\odot}}\xspace}

\newcommand{\Teff}{\ensuremath{T_{e\!f\!f}}\xspace}
\newcommand{\logg}{\ensuremath{\log~g}\xspace}

\newcommand{\mic}{$\mu$m\xspace}
\newcommand{\as}{\hbox{$^{\prime\prime}$}\xspace}

\begin{document}

\title{First light of the VLT planet finder SPHERE}
\subtitle{I. Detection and characterization of the sub-stellar companion GJ\,758\,B\thanks{Based on observations collected at the European Southern Observatory, Chile, during the commissioning of the SPHERE instrument}}

\titlerunning{Detection and characterization of the sub-stellar companion GJ\,758\,B with VLT/SPHERE}

\author{
A. Vigan\inst{1,2} \and 
M. Bonnefoy\inst{3,4} \and
C. Ginski\inst{5} \and
H. Beust\inst{3,4} \and
R. Galicher\inst{6} \and
M. Janson\inst{7,8} \and
J.-L. Baudino\inst{6} \and
E. Buenzli\inst{8} \and
J. Hagelberg\inst{9,10} \and
V. D'Orazi\inst{11,12,13} \and
S. Desidera\inst{11} \and
A.-L. Maire\inst{11} \and
R. Gratton\inst{11} \and
J.-F. Sauvage\inst{14,1} \and
G. Chauvin\inst{3,4} \and
C. Thalmann\inst{15} \and
L. Malo\inst{16} \and
G. Salter\inst{1} \and
A. Zurlo\inst{1,17,18} \and
J. Antichi\inst{11} \and
A. Baruffolo\inst{11} \and
P. Baudoz\inst{6} \and
P. Blanchard\inst{1} \and
A. Boccaletti\inst{6} \and
J.-L. Beuzit\inst{3,4} \and
M. Carle\inst{1} \and
R. Claudi\inst{11} \and
A. Costille\inst{1} \and
A. Delboulb\'e\inst{3,4} \and
K. Dohlen\inst{1} \and
C. Dominik\inst{19} \and
M. Feldt\inst{8} \and
T. Fusco\inst{14,1} \and
L. Gluck\inst{3,4} \and
J. Girard\inst{2,3,4} \and
E. Giro\inst{11} \and
C. Gry\inst{1} \and
T. Henning\inst{8} \and
N. Hubin\inst{20} \and
E. Hugot\inst{1} \and
M. Jacquet\inst{1} \and
M. Kasper\inst{20,3,4} \and
A.-M. Lagrange\inst{3,4} \and
M. Langlois\inst{21,1} \and
D. Le Mignant\inst{1} \and
M. Llored\inst{1} \and
F. Madec\inst{1} \and
P. Martinez\inst{22} \and
D. Mawet\inst{2,23} \and
D. Mesa\inst{11} \and
J. Milli\inst{2,3,4} \and
D. Mouillet\inst{3,4} \and
T. Moulin\inst{3,4} \and
C. Moutou\inst{1,16} \and
A. Orign\'e\inst{1} \and
A. Pavlov\inst{8} \and
D. Perret\inst{6} \and
C. Petit\inst{14} \and
J. Pragt\inst{24} \and
P. Puget\inst{3,4} \and
P. Rabou\inst{3,4} \and
S. Rochat\inst{3,4} \and
R. Roelfsema\inst{24} \and
B. Salasnich\inst{11} \and
H.-M. Schmid\inst{15} \and
A. Sevin\inst{6} \and
A. Smette\inst{2} \and
E. Stadler\inst{3,4} \and
M. Suarez\inst{20} \and
M. Turatto\inst{11} \and
S. Udry\inst{10} \and
F. Vakili\inst{22} \and
Z. Wahhaj\inst{2,1} \and
L. Weber\inst{10} \and
F. Wildi\inst{10} 
}

\institute{
Aix Marseille Universit\'e, CNRS, LAM (Laboratoire d'Astrophysique de Marseille) UMR 7326, 13388, Marseille, France\\ \email{\href{mailto:arthur.vigan@lam.fr}{arthur.vigan@lam.fr}} %1
\and
European Southern Observatory, Alonso de Cordova 3107, Vitacura, Santiago, Chile %2
\and
Universit\'e Grenoble Alpes, IPAG, F-38000 Grenoble, France  %3
\and
CNRS, IPAG, F-38000 Grenoble, France %4
\and
Leiden Observatory, Leiden University, P.O. Box 9513, 2300 RA Leiden, The Netherlands %5
\and
LESIA, Observatoire de Paris, CNRS, Universit\'e Paris Diderot, Universit\'e Pierre et Marie Curie, 5 place Jules Janssen, 92190 Meudon, France %6
\and
Stockholm University, AlbaNova University Center, Stockholm, Sweden %7
\and
Max Planck Institute for Astronomy, K\"onigstuhl 17, 69117 Heidelberg, Germany %8
\and
Institute for Astronomy, University of Hawai'i, 2680 Woodlawn Drive, Honolulu, HI 96822, USA %9
\and
Geneva Observatory, University of Geneva, Chemin des Mailettes 51, 1290 Versoix, Switzerland % 10
\and
INAF - Osservatorio Astronomico di Padova, Vicolo dell'Osservatorio 5, 35122 Padova, Italy %11
\and
Dept. of Physics and Astronomy, Macquarie University, NSW 2109, Sydney, Australia %12
\and
Monash Centre for Astrophysics, Monash University, VIC 3800, Melbourne, Australia %13
\and
ONERA, The French Aerospace Lab BP72, 29 avenue de la Division Leclerc, 92322 Ch\^atillon Cedex, France %14
\and
Institute for Astronomy, ETH Zurich, Wolfgang-Pauli-Strasse 27, 8093 Zurich, Switzerland %15
\and
CNRS, CFHT, 65-1238 Mamalahoa Hwy, Kamuela HI, USA %16
\and
N\'ucleo de Astronom\'ia, Facultad de Ingenier\'ia, Universidad Diego Portales, Av. Ejercito 441, Santiago, Chile %17
\and
Millennium Nucleus ``Protoplanetary Disk'', Departamento de Astronom\'ia, Universidad de Chile, Casilla 36-D, Santiago, Chile %18
\and
University of Amsterdam Sterrenkundig Instituut ``Anton Pannekoek'' Science Park 9041098 XH Amsterdam, the Netherlands %19
\and
European Southern Observatory, Karl-Schwarzschild-Str. 2, 85748 Garching, Germany %20
\and
CRAL, UMR 5574, CNRS, Universit\'e Lyon 1, 9 avenue Charles Andr\'e, 69561 Saint Genis Laval Cedex, France %21
\and
Universit\'e Nice-Sophia Antipolis, CNRS, Observatoire de la C\^ote d'Azur, Laboratoire J.-L. Lagrange, CS 34229, 06304 Nice cedex 4, France %22
\and
California Institute of Technology, 1200 E. California Blvd, Pasadena, CA 91125 %23
\and
NOVA Optical-Infrared Instrumentation Group at ASTRON, Oude Hoogeveensedijk 4, 7991 PD Dwingeloo, The Netherlands %24
}

\date{Received 5 May 2015; accepted 6 November 2015}

\abstract{GJ\,758\,B is a brown dwarf companion to a nearby (15.76~pc) solar-type, metal-rich (${\rm M/H} = +0.2$~dex) main-sequence star (G9V) that was discovered with Subaru/HiCIAO in 2009. From previous studies, it has drawn attention as being the coldest ($\sim$600~K) companion ever directly imaged around a neighboring star. We present new high-contrast data obtained during the commissioning of the SPHERE instrument at the VLT. The data was obtained in $Y$-, $J$-, $H$-, and $K_s$-bands with the dual-band imaging (DBI) mode of IRDIS, providing a broad coverage of the full near-infrared (near-IR) range at higher contrast and better spectral sampling than previously reported. In this new set of high-quality data, we report the re-detection of the companion, as well as the first detection of a new candidate closer-in to the star. We use the new 8 photometric points for an extended comparison of GJ\,758\,B with empirical objects and 4 families of atmospheric models. From comparison to empirical object, we estimate a T8 spectral type, but none of the comparison object can accurately represent the observed near-IR fluxes of GJ\,758\,B. From comparison to atmospheric models, we attribute a $\Teff = 600 \pm 100$~K, but we find that no atmospheric model can adequately fit all the fluxes of GJ\,758\,B. The lack of exploration of metal enrichment in model grids appears as a major limitation that prevents an accurate estimation of the companion physical parameters. The photometry of the new candidate companion is broadly consistent with L-type objects, but a second epoch with improved photometry is necessary to clarify its status. The new astrometry of GJ\,758\,B shows a significant proper motion since the last epoch. We use this result to improve the determination of the orbital characteristics using two fitting approaches, Least-Square Monte Carlo and Markov Chain Monte Carlo. We confirm the high-eccentricity of the orbit (peak at 0.5), and find a most likely semi-major axis of 46.05~AU. We also use our imaging data as well as archival radial velocity data to reject the possibility this is a false positive effect created by an unseen, closer-in companion. Finally, we analyze the sensitivity of our data to additional closer-in companions and reject the possibility of other massive brown dwarf companions down to 4--5~AU.}


\maketitle

\section{Introduction}
\label{sec:introduction}

The direct-imaging search for sub-stellar companions around nearby stars has led to an increasing number of discoveries in the vicinity of our Sun. \object{GJ\,758}\,B \citep{thalmann2009} is one of the brown dwarf companions that stands out of the list. The primary star is a nearby \citep[15.76~pc;][]{vanLeeuwen2007} solar-type (G9V) star, and the inferred effective temperature (\Teff) of GJ\,758\,B is among the lowest ($\sim$600~K) ever recorded for a directly imaged companion. These peculiarities made this system the subject of two separate studies \citep{currie2010,janson2011} in addition to its discovery paper.

Previous observations have provided good spectral coverage of the GJ\,758 system. Common proper motion of the companion with its parent star was determined through two epochs of Subaru/HiCIAO H-band observations detailed in \citet{thalmann2009}. In their work, they also highlight its very low \Teff (550--640~K) and late spectral type (T9). \citet{currie2010} published MMT/Clio $L'$-band measurements of GJ\,758\,B. These data showed the object to have extremely red colors between near- and mid-infrared ($H - L' = 3.29 \pm 0.25$). The latest publication on GJ\,758\,B, by \cite{janson2011}, completed the spectral coverage with measurements in $J$, $H$, CH4S, CH4L, $K_c$, $L'$ and $M_s$ from Subaru/HiCIAO, Gemini/NIRI and Keck/NIRC2. They confirmed again the very low \Teff and late spectral type of the companion, and for the first time they demonstrated the clear methane absorption in $H$-band from the NIRI measurements in the CH4S and CH4L filters. In general, all three papers converged towards a similar picture of a low mass brown dwarf (30--40~\MJup), given the old age of the system (5--9~Gyr). First attempts at an orbit determination for GJ\,758\,B hinted at a large semi-major axis ($30 \leq a \leq 90$~AU) and high eccentricity ($0.4 \leq e \leq 0.7$). However, even though orbital motion is detected, further astrometric monitoring is needed for accurate orbital parameters to be determined. This is due to only having detected a small fraction of the total orbit.

In this work we present new near-infrared (near-IR) photometric data obtained with the SPHERE instrument \citep[Spectro-Polarimetric High-contrast Exoplanet REsearch;][]{beuzit2008}, recently commissioned at the Very Large Telescope (VLT) in Chile. We first present our observations (Sect.~\ref{sec:observations}) and the data reduction and analysis (Sect.~\ref{sec:data_reduction_analysis}). These new observations cover the full near-IR range at much higher contrast than previous observations, allowing us to detect a new candidate companion at closer projected separation than GJ\,758\,B. We provide photometric measurements of GJ\,758\,B with improved sampling and resolution, including the very first measurements of the companion flux in $Y$-band. After revisiting the stellar parameters and age indicators for GJ\,758\,A (Sect.~\ref{sec:stellar_parameters}), we perform an updated modeling of the properties of GJ\,758\,B from comparison to empirical objects and atmospheric models (Sect.~\ref{sec:spectral_analysis}). Finally, we use the new astrometric data point to improve the orbit determination (Sect.~\ref{sec:astrometry_orbital_properties}) before concluding with our sensitivity to additional closer-in companions (Sect.~\ref{sec:sensitivity_additional_companions}).

\section{Observations}
\label{sec:observations}

\begin{table}
  \caption{IRDIS DBI filters wavelength and resolution}
  \label{tab:dbi_filters}
  \centering
  \begin{tabular}{cccc}
  \hline\hline
  Filter pair & Filter & Wavelength & Resolution \\
              &        & (\mic)     &            \\
  \hline
  Y23         & Y2     & 1.022      & 20         \\
              & Y3     & 1.076      & 20         \\
  \hline
  J23         & J2     & 1.190      & 25         \\
              & J3     & 1.273      & 25         \\
  \hline
  H23         & H2     & 1.593      & 30         \\
              & H3     & 1.667      & 30         \\
  \hline
  K12         & K1     & 2.110      & 20         \\
              & K2     & 2.251      & 20         \\
  \hline
  \end{tabular}
\end{table}

\begin{table*}
  \caption{Observing log}
  \label{tab:observing_log}
  \centering
  \begin{tabular}{ccccccccccc}
  \hline\hline
  UT date      & Julian date & Filter pair & DIT\tablefootmark{a} $\times$ NDIT & Dithering\tablefootmark{b}  & T$_{\mathrm{exp}}$ & FoV rot. & Seeing\tablefootmark{c} & Sr\tablefootmark{c} \\
               & (day)       &             & (s)                                &                             & (min)            & (deg)    & (as)                    & (\%)                \\
  \hline                                                                  
  2014-08-13   & 2456882     & J23    & $32 \times 3$                      & 4$\times$4                  & 25.6             & 7.1      & $0.50 \pm 0.06$        & $74 \pm 4$            \\
  2014-08-13   & 2456882     & H23    & $32 \times 3$                      & 4$\times$4                  & 25.6             & 7.2      & $0.58 \pm 0.06$        & $85 \pm 2$            \\
  \hline                                                                  
  2014-08-14   & 2456883     & Y23    & $32 \times 3$                      & 4$\times$4                  & 25.6             & 7.1      & $0.50 \pm 0.11$        & $58 \pm 6$            \\
  2014-08-14   & 2456883     & K12    & $32 \times 3$                      & 4$\times$4                  & 25.6             & 7.2      & $0.44 \pm 0.05$        & $89 \pm 1$            \\
  \hline
  \end{tabular}
  \tablefoot{\tablefoottext{a}{Detector integration time.} \tablefoottext{b}{Detector dithering (see text for details). NDIT images are acquired at each detector dithering positions.} \tablefoottext{c}{The seeing and Strehl ratio estimations are calculated over periods of 10~s every 30~s by the real-time computer. The Strehl ratio is expressed in the mean wavelength of the considered filter pair. The error bar is calculated as the standard deviation of the values.}}
\end{table*}

GJ\,758 was observed as part of the third SPHERE commissioning run in August 2014. The SPHERE planet-finder instrument installed at the VLT \citep{beuzit2008} is a highly specialized instrument, dedicated to high-contrast imaging and spectroscopy of young giant exoplanets. It is based on the SAXO extreme adaptive optics system \citep{fusco2006,petit2014,sauvage2014}, which controls a $41\times41$ actuators deformable mirror, and 4 control loops (fast visible tip-tilt, high-orders, near-infrared differential tip-tilt and pupil stabilization). The common path optics employ several stress polished toric mirrors \citep{hugot2012} to transport the beam to the coronagraphs and scientific instruments. Several types of coronagraphic devices for stellar diffraction suppression are provided, including apodized pupil Lyot coronagraphs \citep{soummer2005} and achromatic four-quadrants phase masks \citep{boccaletti2008}. 

The GJ\,758 observations were acquired with one of the three scientific sub-systems of SPHERE, the infrared dual-band imager and spectrograph \citep[IRDIS;][]{dohlen2008a} in its dual-band imaging mode \citep[DBI;][]{vigan2010} with 4 different filter pairs in the $Y$-, $J$-, $H$- and $K_s$-bands. The spectral characteristics of the filters are provided in Table~\ref{tab:dbi_filters}. The observations were performed in pupil-stabilized mode to perform angular differential imaging \citep[ADI;][]{marois2006} with an apodized pupil Lyot coronagraph \citep{soummer2005} optimized for the $H$-band (ALC\_YJH\_S), which uses a coronagraphic mask of diameter 185 mas. The data were acquired on two consecutive nights, 13$^{\rm th}$ and 14$^{\rm th}$ of August 2014, with a total integration time of $\sim$26~min in each filter pair. The IRDIS detector was dithered on a 4$\times$4 pattern to reduce the effect of the residual flat field noise. At each detector dithering position, a data cube of DIT$\times$NDIT=32$\times$3~s was acquired, resulting in a total of 16 data cubes for each observation. 

The observing sequence in each of the DBI filters was performed as follows:

\begin{itemize}
\item One image of the PSF taken off-axis ($\sim$0.4\as) with the neutral density ND3.5, which reduces the flux by a factor $\sim$3000. The PSF is moved off the coronagraph by applying an offset on the near-IR differential tip-tilt plate. During this observation, the AO visible tip-tilt and high-order loops remain closed to provide a diffraction-limited PSF;
\item A ``star center'' coronagraphic image where four symmetric satellite spots are created by introducing a periodic modulation on the deformable mirror. This data is used in subsequent analysis to determine an accurate position of the star center behind the coronagraph, and hence the center of field rotation;
\item The coronagraphic sequence as previously described;
\item An additional off-axis PSF to evaluate the variations of the observing conditions between the beginning and end of the sequence.
\end{itemize} 

For commissioning purposes, data from the SPARTA real-time computer of the SAXO extreme AO system \citep{fusco2014} were collected at regular intervals in parallel of all the observations. This includes in particular images from the differential tip-tilt sensor (DTTS). This sensor removes a minute fraction of the incoming flux in the near-IR arm (at 1.6~\mic) to image the PSF just before the coronagraph, and uses it as input for the DTTS loop that maintains the PSF locked on the coronagraph once the observing sequence has started. Every 30 seconds, the 30 seconds average of the non-coronagraphic PSF on the DTTS is saved in the SPARTA files, allowing a fine monitoring of the PSF motion and flux variation at the level of the coronagraph. 

Standard calibrations for the DBI mode were acquired in the morning as part of the IRDIS calibration plan. Instrumental backgrounds were taken for both the coronagraphic and off-axis exposures with proper DIT values. Detector flat fields were also acquired in each of the DBI filter pairs.

\section{Data reduction \& analysis}
\label{sec:data_reduction_analysis}

\begin{figure*}
  \centering
  \includegraphics[width=1.0\textwidth]{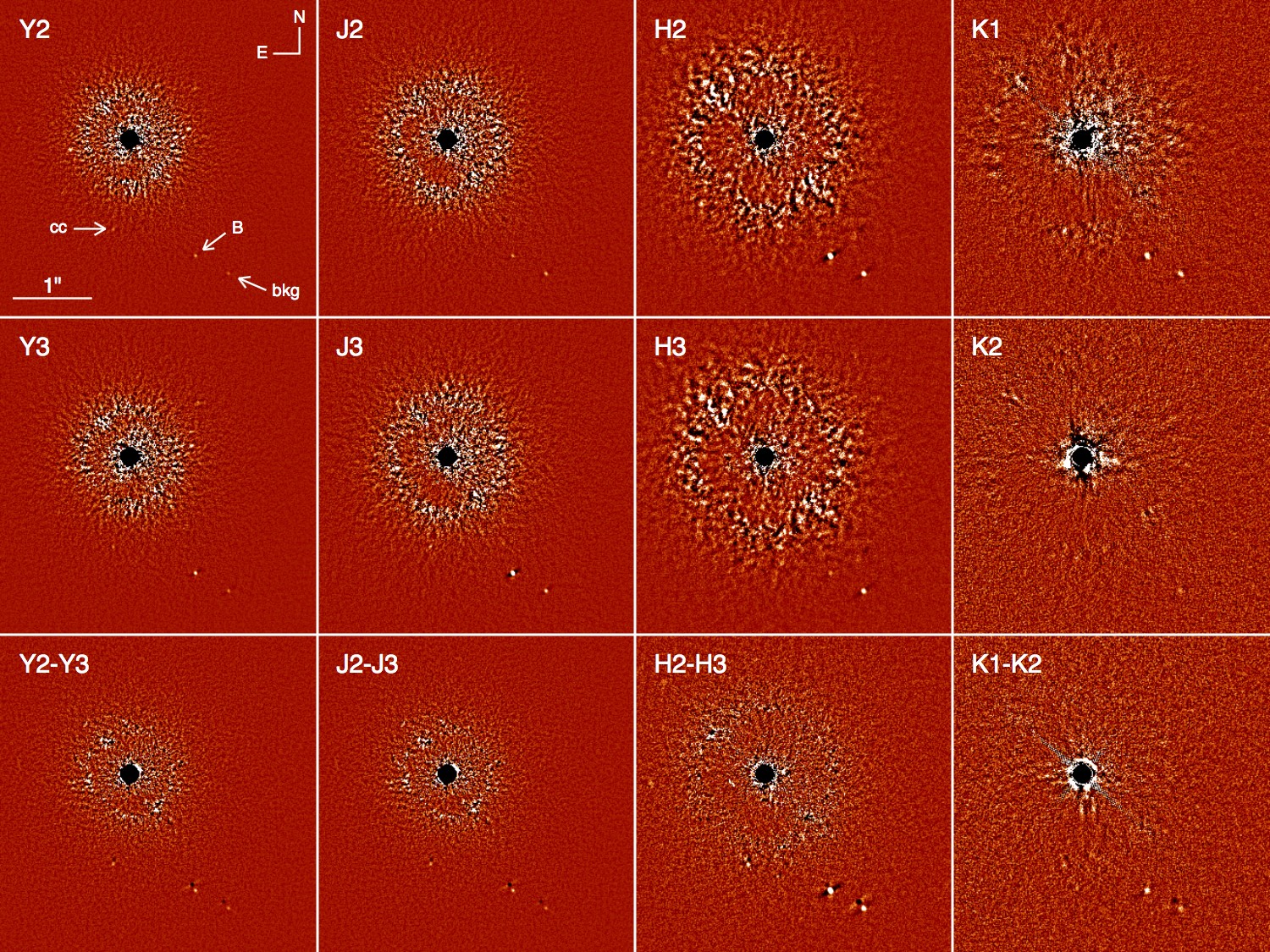}
  \caption{Images of GJ\,758 after ADI and SDI processing in all IRDIS DBI filters. For each filter pair, the top and middle rows present the ADI analysis of the data in the first and second filters respectively, and the bottom row presents the result of the SDI+ADI analysis. For the ADI analysis, 5 PCA modes were subtracted, while for the SDI+ADI analysis, only a single mode was subtracted. Three objects are clearly identified in the data: GJ\,758\,B (``B''), a background star (``bkg'') and a new candidate companion (``cc''). The spatial and display scales are identical between all images. The SDI images display the characteristic negative/positive pattern expected for physical objects that present flux in both DBI filters. For the highly methane-bearing object GJ\,758\,B, the flux difference between the H2 and H3 filters is clearly visible.}
  \label{fig:GJ758_all_filters}
\end{figure*}

\begin{table}
  \caption{Mean plate scale measured from observations of the 47\,Tuc globular cluster.}
  \label{tab:astrometric_calibration}
  \centering
  \begin{tabular}{cc}
  \hline\hline
  Filter  & Plate scale      \\
          & (mas/pixel)      \\
  \hline
  Y2      & $12.287 \pm 0.006$ \\
  Y3      & $12.282 \pm 0.006$ \\
  \hline
  J2      & $12.267 \pm 0.006$ \\
  J3      & $12.262 \pm 0.006$ \\
  \hline
  H2      & $12.263 \pm 0.006$ \\
  H3      & $12.258 \pm 0.006$ \\
  \hline
  \end{tabular}
\end{table}

\begin{table*}
  \caption{Astrometry and photometry of GJ\,758\,B and the newly detected candidate relative to primary}
  \label{tab:astrophotometry_gj758}
  \centering
  \begin{tabular}{cccccc}
  \hline\hline
  Filter & $\Delta\alpha$ & $\Delta\delta$ & Sep.      & P.A.              & $\Delta$mag \\
         & (mas)          & (mas)          & (mas)     & (deg)             & (mag)       \\
  \hline
  \multicolumn{6}{c}{GJ\,758\,B} \\
  \hline
  Y2     & $-793 \pm 4 $& $-1501 \pm 3 $& $1698 \pm 3 $& $207.85 \pm 0.13$ & $14.90 \pm 0.19$ \\
  Y3     & $-789 \pm 4 $& $-1501 \pm 2 $& $1698 \pm 2 $& $207.86 \pm 0.12$ & $14.20 \pm 0.09$ \\
  J2     & $-789 \pm 4 $& $-1499 \pm 4 $& $1694 \pm 4 $& $207.77 \pm 0.16$ & $14.97 \pm 0.24$ \\
  J3     & $-789 \pm 4 $& $-1499 \pm 2 $& $1694 \pm 2 $& $207.78 \pm 0.12$ & $12.89 \pm 0.17$ \\
  H2     & $-791 \pm 4 $& $-1501 \pm 2 $& $1697 \pm 2 $& $207.80 \pm 0.12$ & $12.95 \pm 0.11$ \\
  H3     & $-792 \pm 7 $& $-1500 \pm 7 $& $1696 \pm 7 $& $207.83 \pm 0.24$ & $15.29 \pm 0.41$ \\
  K1     & $-784 \pm 7 $& $-1500 \pm 8 $& $1692 \pm 8 $& $207.58 \pm 0.25$ & $13.46 \pm 0.20$ \\
  K2     & $-794 \pm 12$& $-1508 \pm 12$& $1704 \pm 12$& $207.76 \pm 0.50$ & $14.21 \pm 0.34$ \\
  \hline
  \multicolumn{6}{c}{New candidate companion} \\
  \hline
  Y2     &  $239 \pm 6 $& $-1132 \pm 12$& $1156 \pm 12$& $168.09 \pm 0.31$ & $15.45 \pm 0.62$ \\
  Y3     &  $241 \pm 7 $& $-1133 \pm 11$& $1158 \pm 11$& $168.00 \pm 0.35$ & $14.40 \pm 0.86$ \\
  J2     &  $242 \pm 6 $& $-1141 \pm 12$& $1167 \pm 12$& $168.04 \pm 0.31$ & $14.67 \pm 0.60$ \\
  J3     &  $244 \pm 10$& $-1137 \pm 19$& $1163 \pm 18$& $167.89 \pm 0.51$ & $15.01 \pm 0.92$ \\
  H2\tablefootmark{a} &  $240 \pm 8$ & $-1139 \pm 9$ & $1164 \pm 9$ & $168.11 \pm 0.40$ & $14.67 \pm 0.61$ \\
  H3\tablefootmark{a} & $\ldots$     & $\ldots$      & $\ldots$     & $\ldots$          & $14.64 \pm 0.61$ \\
  K1     &  $263 \pm 13$& $-1157 \pm 20$& $1187 \pm 19$& $167.18 \pm 0.65$ & $13.77 \pm 0.30$ \\
  \hline
  \end{tabular}
  \tablefoot{\tablefoottext{a}{The H23 photometry and astrometry of the new candidate companion are determined jointly using a combination of SDI and ADI (see text for details). In this context, the astrometry is only relevant in the H2 filter.}}
\end{table*}

The data were analyzed with two separate pipelines, which are described in this section. 

The LAM-ADI pipeline is similar to that described in \citet{vigan2012} after updates to work with the SPHERE/IRDIS data. The calibrations (backgrounds, flat) were created using the preliminary release (v0.14.0-2) of the SPHERE data reduction and handling (DRH) software \citep{pavlov2008}. Each of the images in the coronagraphic observing sequences were background subtracted and divided by the flat field in the appropriate DBI filters. Bad pixels were corrected using bad pixel maps created with the DRH by replacing them with the median of neighboring good pixels. Finally, all images were aligned to a common center using the star center data acquired at the beginning of the sequence. For this purpose, the 4 satellite spots inside the AO control radius were fitted with a 2D Gaussian function using the \texttt{MPFIT} non-linear least squares curve fitting software \citep{markwardt2009}. The accuracy of the centering using this procedure has been determined to be better than 0.1 pixel ($\sim$1.2~mas) for bright stars during the first SPHERE commissioning run in May 2014. For the recentering of the science frames, the shift introduced by the detector dithering procedure was also taken into account, and the 0.06 pixel (0.74~mas) accuracy of the dithering motion stage was included into the astrometric error budget. For each filter pair, the calibration process was applied independently to each of the two wavelengths acquired simultaneously with IRDIS, resulting in two separate pre-processed ADI data cubes.

The ADI data cubes were processed with the LAM-ADI pipeline using a principal component analysis (PCA) implementation following the KLIP approach \citep{soummer2010}. The number of subtracted modes, minimum and maximum radii for the analysis were varied over a wide range, but the companion was recovered in all analyses. Figure~\ref{fig:GJ758_all_filters} shows the signal of the companion in all of the IRDIS DBI filters. The companion is recovered in all filters with a signal-to-noise ratio (SNR) greater than 6, except in the K2 filter where it is only marginally detected with an SNR of $\sim$2.5. As already presented in \citet{janson2011}, the companion displays a clear methane absorption in $H$-band with a flux about 9 times fainter in H3 than in H2. Images were also processed using a combination of spectral differential imaging \citep[SDI;][]{racine1999} and ADI to attenuate even more the speckle noise and look for additional fainter candidates. In addition to the detection of GJ\,758\,B, we report the re-detection of the background star already identified by \citet{janson2011}, and the detection of a new candidate located $\sim$1.1\as South of the star in all filters except K2. Although not directly detectable in H2 and H3 with ADI only, the candidate was easily identifiable in the SDI+ADI processed image.

The precise astrometry and photometry of the companion and new candidate was estimated using ``negative fake companion'' subtraction in the pre-processed ADI data cubes \citep{marois2010}. A rough estimation of the object position and contrast is first performed using a 2D Gaussian fit. Then these initial guesses are used as a starting point for a Levenberg-Marquardt least-squares minimization routine where the position and contrast of the negative fake companion are varied to minimize the residual noise after ADI-processing in a circular aperture of radius $\lambda/D$ centered on the position of the companion. When a minimum is reached, the position and contrast of the fake companion are taken as the optimal values for the astrometry and photometry. Note that this procedure is also applicable for analyses combining SDI and ADI, by minimizing the residuals in an aperture that covers the position of the companion in the first filter, and in the second filter after spatial rescaling. The error bars for the fitting process are then calculated by varying the position and contrast of the fake companion until the variation of the reduced $\chi^{2}$ reaches a level of 1$\sigma$. 

The data were analyzed independently with the LESIA pipeline for a cross-check of the astrometry and photometry. This pipeline uses a similar approach for the pre-processing of the ADI data cubes, but for the speckles subtraction it uses an upgrade of the Template Locally Optimized Combination of Images  (TLOCI) algorithm derived from the one presented in \citet{marois2014}. Only ADI is used (no SDI) to avoid issues with the photometry calibration \citep{maire2014}. Hence, for each dual-band filter sequence, it calibrates the speckle pattern in each individual frame, rotates the frames to align North up, and median-combines all the frames to obtain the final image. To derive the photometry and the astrometry of the detected sources, the pipeline uses the unsaturated PSF of the central star (recorded before and after the coronagraphic sequence) to build a data cube composed of fake companions at the positions of the detected sources accounting for the field-of-view rotation in each frame and smearing during exposures. Then, the frames of this data cube are combined using the TLOCI coefficients that were used to obtain the image where the point-source was detected. The resulting frames are aligned in the same way as the science data to obtain an image that gives a model of the off-axis sources in the TLOCI images at the positions of the detections, accounting for TLOCI self-subtraction and distortions. Finally, the sub-pixel position and the flux of the modeled images are adjusted to optimize the subtraction of the model to the real image within a 1.5~$\lambda/D$-radius disk centered on the detection \citep{galicher2011}. The error bars account for the variations of the stellar flux during the sequence (estimated from the global speckle intensity variations), and the accuracy of the fitting of the companion image models to the real images.

For calibrating the distortion, plate scale, and orientation of the IRDIS images, a field in the outer regions of the \object{47\,Tuc} globular cluster was observed in different instrumental configurations (Maire et al., A\&A, in press). The 47\,Tuc field was selected because it includes a bright star for adaptive optics guiding and was accurately calibrated using \textit{Hubble Space Telescope} (HST) observations \citep{bellini2014}. The plate scales for the different DBI filters are summarized in Table~\ref{tab:astrometric_calibration}. Since it was not calibrated in the K12 filter pair during the commissioning, we assumed the same value as for the H23 filter pair. The true North correction measured for this commissioning run is $-1.636 \pm 0.013\deg$, and the correction of the orientation also takes into account the zero point orientation of the derotator in pupil-stabilized mode, which was measured to be $135.87 \pm 0.03\deg$.

Relative photometry and astrometry of the companion and the newly detected candidate are reported in Table~\ref{tab:astrophotometry_gj758}. Both pipelines agree within their respective error bars. The values reported in the table correspond to the average of the results from both pipelines, and the respective errors bars have been quadratically added. The final error bars for the photometry include the fitting error detailed above, the variation of the non-coronagraphic PSF measured on the DTTS images (see Sect.~\ref{sec:observations}), and the level of speckle residuals estimated at the same separation as the detections. The astrometric error bars include the fitting error, and the uncertainties on the star center, dithering motion, plate scale, derotator zero point and true North correction. Note that for astrometry, the reference values are those from the H23 filter pair, which has been the most accurately calibrated. 

\section{Stellar parameters}
\label{sec:stellar_parameters}

A reassessment of stellar parameters of GJ\,758 is warranted considering their relevance in the derivation of the properties of its sub-stellar companion and to explain its peculiar features discussed in Sect.~\ref{sec:spectral_analysis}. GJ\,758 is classified as an old star \citep[age 0.7--8.7~Gyr; ][]{janson2011}, and we revisit here the various age indicators, following the procedures and calibrations described in \cite{2015A&A...573A.126D}, as well as the chemical composition.

\subsection{Kinematic parameters}

Adopting the trigonometric parallax, the proper motion and errors bars by \cite{vanLeeuwen2007}, and the absolute radial velocity by \citet{2002ApJS..141..503N} with an error of 0.50~km/s, space velocities U,V,W = $-21.1 \pm 0.2$; $-14.1 \pm 0.5$; $-3.0 \pm 0.2$~km/s are obtained. These are very similar to those of the Argus association (U,V,W = $-21.5 \pm 0.9$,$-12.2 \pm 1.7$,$-4.6 \pm 2.7$). Although the \texttt{BANYAN II} on-line tool \citep{2014ApJ...783..121G} yields a membership probability of 97.8\%, which would correspond to a very young age of 40~Myr, the full version of the \texttt{BANYAN I} tool \citep{2013ApJ...762...88M}, which takes into account both kinematic and photometric informations, yields a 100\% probability to the hypothesis that GJ\,758 is an old field star.

\subsection{Abundance analysis}
\label{sec:abu}

We determined spectroscopic stellar parameters and chemical abundances for GJ\,758 by exploiting a high-resolution (R=42,000), high signal-to-noise (SNR=164 at 5500~\AA) ELODIE spectrum\footnote{\url{http://atlas.obs-hp.fr/elodie/fE.cgi?c=o\&o=GJ758}}, which provides a wavelength coverage from 3850~\AA~to 6800~\AA. The spectrum was downloaded from the on-line ELODIE archive \citep{moultaka14}, which provides reduced data products. This investigation aims at chemically {\it tagging} our target star, in order to ascertain whether the abundance pattern is compatible with the Argus association, whose chemical composition has been recently presented by \cite{desilva2013}. Argus reflects a roughly solar chemical composition, with ${\rm[Fe/H]} = -0.06 \pm 0.05$~dex and all [X/Fe] ratios within 0.15 dex from the solar values, with the notable exception of barium (see discussion below). 

We carried out a homogeneous and strictly differential analysis for GJ\,758 with respect to Argus members published in that previous work, by utilizing the same code (MOOG by \citealt{sneden73}, 2014 version), line lists, techniques, and grid of model atmospheres (\citealt{kur93}, solar-scaled models and no convective overshooting). Effective temperature (\Teff) and surface gravity (log~$g$) were derived by imposing excitation and ionisation equilibrium, that is no spurious trend of A(Fe) with the excitation potentials of the spectral features and agreement (within 0.05~dex) of iron abundances from Fe~{\sc i} and Fe~{\sc ii}, respectively. The microturbulence velocity ($\xi$) was instead calculated requiring that abundances from Fe~{\sc i} lines show no trend with reduced equivalent widths. We performed equivalent width analysis for Fe, Na, Mg, Al, Si, Ca, Ti, Cr, and Ni, whereas the Ba abundance was inferred via spectral synthesis, including hyperfine structure and isotopic splitting, as in \cite{desilva2013}.

Internal (random) uncertainties affecting our derived abundances were computed in the standard way, that is by adding in quadrature errors due to the equivalent-width (EW) measurements (or to the best-fit determination in the case of spectral synthesis) and those related to the adopted set of atmospheric parameters (\Teff, \logg, and $\xi$). The total internal errors for [Fe/H] as well as for [X/Fe] ratios are given in Table~\ref{tab:abundances} (we refer the reader to \citealt{desilva2013} for further details on the error budget calculation).

\begin{table}
\caption{Spectroscopic stellar parameters and abundances for GJ\,758.}\label{tab:abundances}
\centering
\begin{tabular}{lc}
\hline\hline
\Teff (K)              & $5498 \pm 50$ \\
\logg (cm~s$^{-2}$)    & $4.53 \pm 0.10$ \\ 
$\xi$ (km~s$^{-1}$)    & $1.12 \pm 0.10$ \\
${\rm [Fe/H]}${\sc i}  & $0.18 \pm 0.05$ \\
${\rm[Fe/H]}${\sc ii}  & $0.13 \pm 0.08$ \\
${\rm[Na/Fe]}$         & $0.12 \pm 0.05$ \\
${\rm[Mg/Fe]}$         & $0.11 \pm 0.05$ \\
${\rm[Al/Fe]}$         & $0.12 \pm 0.05$ \\
${\rm[Si/Fe]}$         & $0.01 \pm 0.05$ \\
${\rm[Ca/Fe]}$         & $0.03 \pm 0.03$ \\
${\rm[Ti/Fe]}${\sc i}  & $0.09 \pm 0.05$ \\
${\rm[Ti/Fe]}${\sc ii} & $0.07 \pm 0.08$ \\
${\rm[Cr/Fe]}${\sc i}  & $0.03 \pm 0.05$ \\
${\rm[Cr/Fe]}${\sc ii} & $0.07 \pm 0.06$ \\
${\rm[Ni/Fe]}$         & $0.04 \pm 0.03$ \\
${\rm[Ba/Fe]}${\sc ii} & $0.00 \pm 0.12$ \\
\hline\hline
\end{tabular}
\end{table}

\begin{figure}
\centering
\includegraphics[width=\columnwidth]{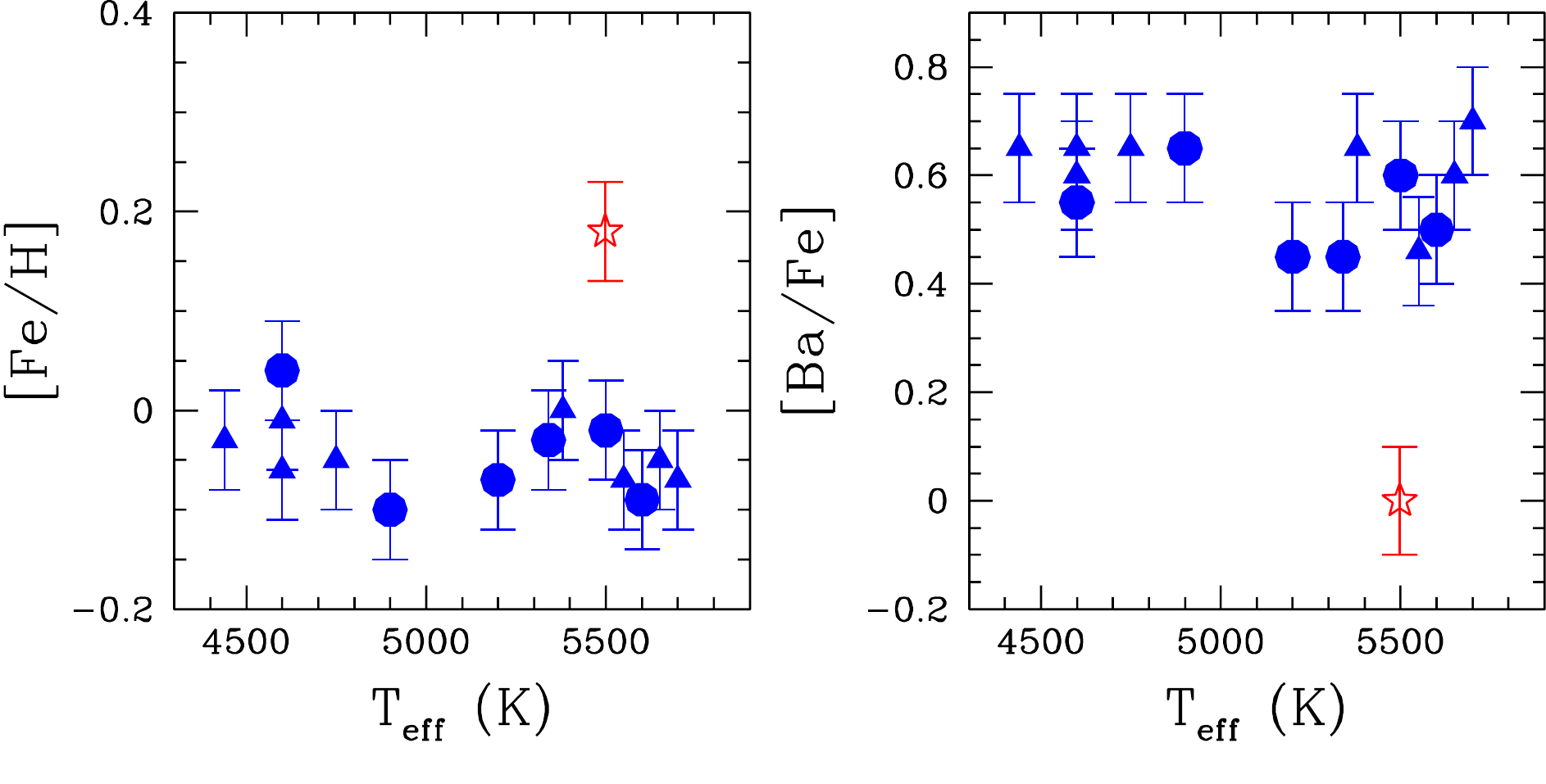}
\caption{Fe and Ba abundances $versus$ effective temperatures for GJ\,758 (starred symbol), the Argus association and the open cluster IC\,2391 (triangles and circles, respectively, from \citealt{desilva2013}).}\label{fig:abu}
\end{figure}

We found a metallicity of ${\rm[Fe/H]} = 0.18 \pm 0.05$, which agrees very well with previous determinations by e.g., \cite{soubiran2008}, \cite{takeda07}, and \cite{maldonado12} and points to super-solar heavy elements abundances for this star. The abundances of $\alpha$-elements Si and Ca as well as the Fe-peak Cr and Ni match a solar-scaled pattern, whereas Na, Mg, Al, and Ti (though to a less extent) seem to exhibit a modest enhancement, but still consistent with solar abundances within the observational uncertainties. The metallicity distribution as a function of effective temperatures is shown in the left-hand panel of Fig.~\ref{fig:abu}: we report [Fe/H] values for GJ\,758 along with stars belonging to Argus (filled circles) and to the open cluster \object{IC\,2391} (triangles), deemed to share a common origin with the young association. It is quite clear from Fig.~\ref{fig:abu} that GJ\,758 stands itself out from the cluster/association distribution, being [Fe/H] roughly $\sim$0.25~dex higher.

Barium deserves a brief, separate discussion. Firstly identified by \cite{dorazi09}, and subsequently confirmed by several studies (e.g., \citealt{yong12}; \citealt{jacobson13}; \citealt{mishenina13}), the Ba abundance shows a decreasing trend with the open cluster's age. The younger the cluster, the higher its Ba content. The reason of such peculiar and unique pattern is still matter of debate: it has been suggested that the efficiency in the production of the s-process elements in low-mass AGB stars is higher than what is predicted form standard stellar evolution models and input physics has to be revised \citep{dorazi2009,maiorca12}. However, subsequent investigations have shown that the picture might not be that straightforward. The fact that the Ba overabundance is not accompanied by a similar behaviour in other s-process elements (e.g., Y, La) makes unlikely this explanation. We refer the reader to \citealt{dorazi12} for a wider discussion of this topic. Regardless of the nature of the super-solar Ba content, [Ba/Fe] ratios range from extremely high values of approximately $\sim$0.6~dex for pre-main sequence clusters, such as e.g., \object{IC\,2602} and IC\,2391 (\citealt{dr09}) to solar values, or even lower, for clusters a few Gyr old. \cite{desilva2013} corroborated this observational evidence and obtained a mean abundance of ${\rm[Ba/Fe]} = 0.53 \pm 0.03$ (rms=0.08~dex) for the Argus association and ${\rm[Ba/Fe]} = 0.62 \pm 0.02$ (rms=0.07) for IC\,2391 (see the right-hand panel of Fig.~\ref{fig:abu}). Conversely, we gathered a ${\rm[Ba/Fe]} = 0.00 \pm 0.12$ for our star, which implies a difference in the Ba content more than a factor of 3.5. Thus, in terms of chemical composition, Ba provides us with the strongest observational constraint: GJ\,758 cannot be born from the same molecular cloud as Argus.

\subsection{Age indicators}

GJ\,758 is known to have a low activity level as resulting from several measurements in the literature: $\log R_{HK} = -4.94$ \citep{2004ApJS..152..261W}, -5.015 \citep{2010ApJ...725..875I}; -5.060 \citep{1991ApJS...76..383D,2008ApJ...687.1264M}. The calibration by \citet{2008ApJ...687.1264M} yields values of 5.5-7.7 Gyr for these activity values. The availability of multi-epoch measurements of chromospheric activity spanning several years indicate that this is not the result of a poor sampling of an activity cycle. The X-ray non-detection in the ROSAT All Sky Survey \citep{1999A&A...349..389V,2000IAUC.7432R...1V} (which would imply $\log L_{X}/L_{bol} < -5.8$ and then an age > 3~Gyr), the small projected rotational velocity (0--2~km/s) and the small photometric variability (0.008~mag from \emph{Hipparcos}) further support the low activity level of GJ\,758, as expected for a few-Gyr old star. 

Lithium is another highly-sensitive age indicator for young stars. From the analysis of the spectrum described in Sect.~\ref{sec:abu} the Li 6708\AA~ resonance line is not detected, confirming the null result by \citet{2005PASJ...57...45T}. For stars with GJ\,758 colors, detectable amounts of lithium vanish at about the age of the Hyades. Therefore, the lack of lithium allows us to infer a stellar age older than 600~Myr.
 
While stellar members of young moving groups display significant scatter in the age indicators \citep[see ][for the case of Argus]{2011A&A...529A..54D}, we are not aware of late G-type stars which are confirmed members of young moving groups which have such a low activity level and lack of lithium. The analysis of these indicators therefore converges with the chemical tagging in ruling out Argus membership for GJ\,758.

Using the spectroscopic effective temperature and metallicity, and the \emph{Hipparcos} $V$ magnitude and trigonometric parallax, we derive age and masses from isochrone using the PARAM interface \citep{2006A&A...458..609D} \footnote{\url{http://stev.oapd.inaf.it/cgi-bin/param}} and the stellar models by \citet{2012MNRAS.427..127B}. Limiting possible input values to age larger than 0.6~Gyr, as resulting from the lack of lithium, the resulting age is $2.2 \pm 1.4$~Gyr and the stellar mass $0.97 \pm 0.02$~\MSun.

\subsection{Summary}

All age indicators indicate that GJ\,758 is an old star, with lithium providing a tight lower limit at 600 Myr. Chemical tagging derived from an homogeneous comparison of abundances of several elements with those of confirmed members of Argus association and IC\,2391 open cluster also rules out a link between GJ\,758 and Argus, with Barium abundance suggesting an age similar to the Sun. Therefore, we conclude that the kinematic parameters of GJ\,758 are similar to those of Argus association just by chance, confirming the statistical nature of kinematic ages and the need for independent youth indications to conclusively infer membership in young moving groups \citep{2014ApJ...783..121G,2015A&A...573A.126D}. The young-disk kinematics decrease the probability of a star significantly older than the Sun. The age of the system is likely within 1 to~6 Gyr, and the most probable value around 3~Gyr, with isochrone fitting yielding younger values than chromospheric activity. We also confirm the moderate super-solar metallicity of the star.

\section{Spectro-photometric analysis}
\label{sec:spectral_analysis}

The new SPHERE photometry is complementary to the existing set of photometric data points on the spectral energy distribution (SED) of the companion obtained by \cite{janson2011}. In the following we use the more complete SED to refine the properties of GJ\,758\,B.

\subsection{Fluxes and magnitudes}

\begin{figure}
\begin{center}
 \includegraphics[width=\columnwidth]{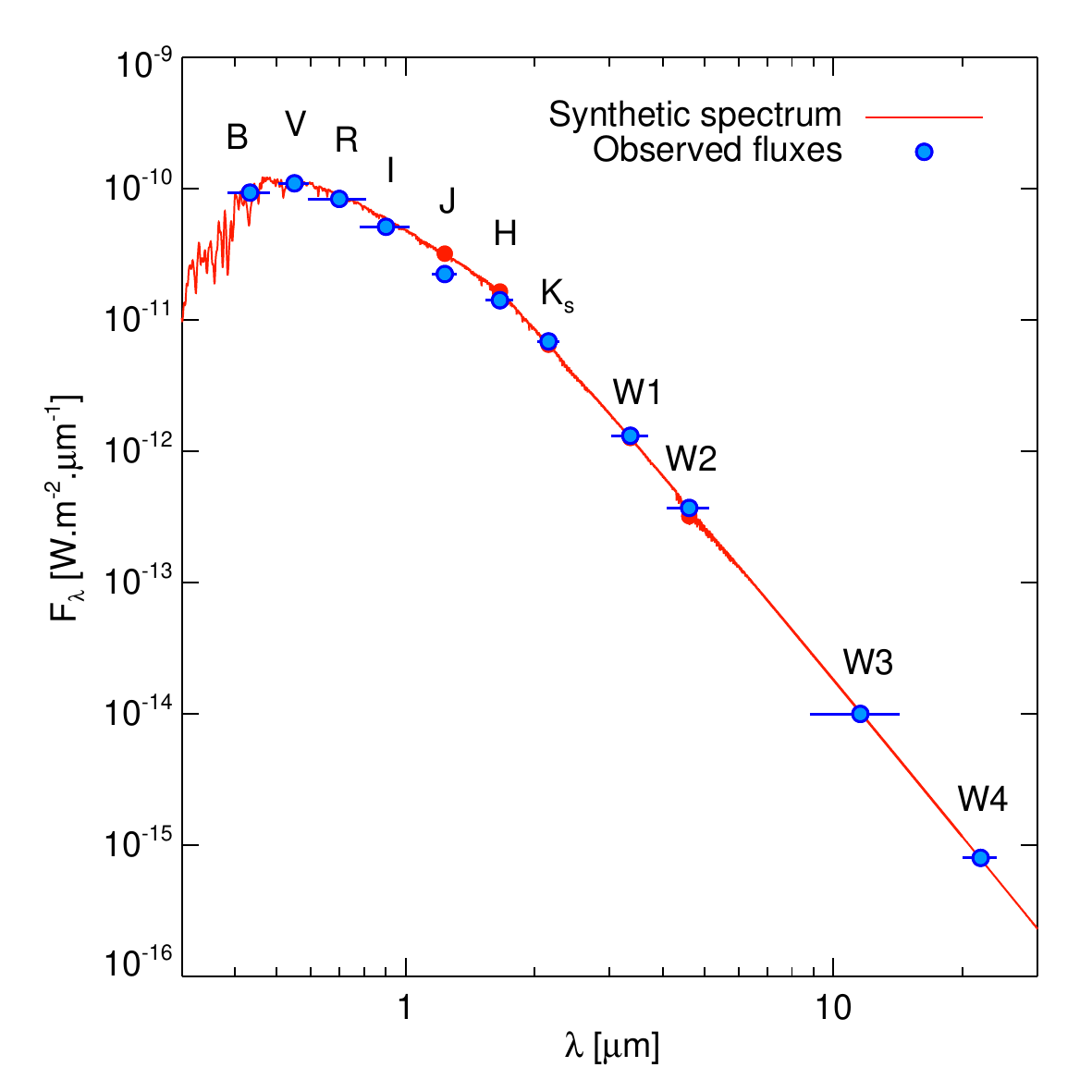}
  \caption{GAIA-COND synthetic spectrum adjusted onto the spectral energy distribution of GJ\,758\,A built from a compilation of optical, near-infrared, and mid-infrared photometry. The 2MASS J, H, K$_{s}$ and WISE W1-W2 photometry data were excluded from the fit because the star was saturating in the 2MASS images.}
\label{Fig:FigSEDstar}
\end{center}
\end{figure}

We retrieved the apparent fluxes corresponding to the SPHERE/IRDIS  photometry of the companion using  the contrast ratio listed in Table~\ref{tab:astrophotometry_gj758} and following a three-steps process: 

\begin{itemize}
\item We first built the  0.4-22.1 $\mu$m SED of the star from the  Tycho $B_{T}$, $V_{T}$ \citep{1997A&A...323L..57H}, USNO-B $R$, and $I$ \citep{2003AJ....125..984M}, and WISE W3-W4 photometry \citep{2013yCat.2328....0C}. The 2MASS $J$, $H$, $K_{s}$ \citep{2003tmc..book.....C} and W1-W2 photometry could not be used because of  the saturation of the star \citep[see ][]{janson2011}. The optical photometry was converted to apparent fluxes using the Gemini flux-conversion tool\footnote{\url{http://www.gemini.edu/sciops/instruments/midir-resources/imaging-calibrations/fluxmagnitude-conversion}}. We considered the WISE zero points reported in \cite{2011ApJ...735..112J} for the infrared part. 
\item We adjusted a GAIA-COND model \citep{2005ESASP.576..565B} with $\Teff = 5400$~K, $\logg = 4.0$~dex, and ${\rm M/H} = 0.0$ onto GJ\,758\,A fluxes values. This model has atmospheric parameters close to the ones determined from  high-resolution spectra of the star \citep[$\Teff = 5435$~K, $\logg = 4.0$, ${\rm M/H} = 0.12$; ][]{2004A&A...427..933K}. The GAIA model reproduces well the SED of GJ\,758 (Fig.~\ref{Fig:FigSEDstar}), including the 2MASS $K_{s}$ band photometry, which appears to be less affected by the saturation. 
\item We derived the mean stellar flux into the SPHERE/IRDIS passbands using the flux-calibrated GAIA spectrum, and the tabulated  filter widths reported in Table~\ref{tab:dbi_filters}. 
\end{itemize} 

The remaining fluxes of GJ\,758\,B  were estimated directly from a flux-calibrated spectrum of Vega \citep{2007ASPC..364..315B}, the Keck/NIRC2 and Gemini/NIRI magnitudes of the companions reported in \cite{janson2011}, and corresponding filter transmission curves. The effect of the telluric absorption on the final flux estimates for the companion was simulated using the ESO sky model calculator\footnote{\url{https://www.eso.org/observing/etc/bin/gen/form?INS.MODE=swspectr+INS.NAME=SKYCALC}} \citep{noll2012,jones2013}. We considered two altitudes of targets above the horizon (90 and 30$^{\circ}$) to simulate dry and wet conditions. The effect is found to be negligible compared to the error on the companion photometry. The final estimated fluxes of  GJ\,758\,B considered for the following analysis are reported in Table \ref{Tab:Fluxesb}. The fluxes in the overlapping narrow-band K1 and $K_{c}$ filters are almost identical. This is an indication that our flux-conversion methods yield consistent results. 

\begin{table}[t]
\caption{Apparent fluxes of GJ\,758\,B.}
\label{Tab:Fluxesb}
\centering
\begin{tabular}{cccccc}
\hline \hline 
Filter &$\lambda$ &$\Delta \lambda$ & $F_{\lambda}$ &  $\Delta F_{\lambda}$  & Ref \\
  &(nm) &(nm) &(W.m$^{-2}.\mu$m$^{-1}$) &(W.m$^{-2}.\mu$m$^{-1}$) & \\
\hline
Y2               & 1022 & 51    & 5.074$\times 10^{-17}$       & 9.703$\times 10^{-18}$ & 1 \\
Y3               & 1076 & 54    & 8.526$\times 10^{-17}$       & 7.368$\times 10^{-18}$ & 1 \\	
J2               & 1190 & 48    & 3.449$\times 10^{-17}$       & 8.532$\times 10^{-18}$ & 1 \\
J                & 1250 & 180   & 1.126$\times 10^{-16}$       & 2.280$\times 10^{-17}$ & 2 \\
J3               & 1273 & 51    & 2.070$\times 10^{-16}$       & 3.508$\times 10^{-17}$ & 1 \\			
CH4S             & 1580 & 103   & 4.074$\times 10^{-17}$       & 8.240$\times 10^{-18}$ & 2 \\
H2               & 1593 & 53    & 1.177$\times 10^{-16}$       & 1.255$\times 10^{-17}$ & 1 \\
H                & 1650 & 290   & 2.536$\times 10^{-17}$       & 5.130$\times 10^{-18}$ & 2 \\
H3               & 1667 & 56    & 1.222$\times 10^{-17}$       & 5.605$\times 10^{-18}$ & 1 \\
CH4L             & 1690 & 110   & $\leq$1.457$\times 10^{-17}$ & \dots                & 2 \\
$\mathrm{K_{c}}$ & 2098 & 28    & 2.747$\times 10^{-17}$       & 5.550$\times 10^{-18}$ & 2 \\
K1               & 2110 & 105   & 2.897$\times 10^{-17}$       & 5.859$\times 10^{-18}$ & 1 \\
K2               & 2251 & 112   & 1.145$\times 10^{-17}$       & 4.209$\times 10^{-18}$ & 1 \\
L'               & 3776 & 700   & 2.163$\times 10^{-17}$       & 2.090$\times 10^{-18}$ & 2 \\
$\mathrm{M_{s}}$ & 4670 & 241   & $\leq$5.257$\times 10^{-17}$ & \dots                 & 2 \\
\hline
\end{tabular}
\tablebib{(1) This work; (2) \citet{janson2011}.}
\end{table}
 
\begin{table}[t]
\caption{Absolute magnitudes of GJ\,758\,A, GJ\,758\,B, and of the candidate companion estimated from the contrast ratio and the model spectrum of the star.}
\label{Tab:MagAb}
\centering
\begin{tabular}{cccc}
\hline \hline 
Filter & GJ\,758\,A\tablefootmark{a} & GJ\,758\,B & New c.c.   \\
\hline
Y2 & $4.29\pm0.03$ & $19.19\pm0.20$ & $19.74\pm0.62$ \\
Y3 & $4.23\pm0.03$ & $18.43\pm0.10$ & $18.73\pm0.86$ \\
J2 & $4.09\pm0.03$ & $19.06\pm0.25$ & $18.76\pm0.60$ \\
J3 & $3.94\pm0.03$ & $16.83\pm0.18$ & $18.95\pm0.92$ \\
H2 & $3.64\pm0.03$ & $16.59\pm0.12$ & $18.31\pm0.61$ \\
H3 & $3.59\pm0.03$ & $18.88\pm0.42$ & $18.23\pm0.61$ \\
K1 & $3.58\pm0.03$ & $17.03\pm0.21$ & $17.35\pm0.62$ \\
K2 & $3.57\pm0.03$ & $17.78\pm0.35$ & \dots          \\
\hline
\end{tabular}
\tablefoot{\tablefoottext{a}{0.03 mag uncertainty assumed based on the SED fit of GJ\,758\,A and on the error on the available optical+WISE photometry of the star.}}
\end{table}

\subsection{Comparison of GJ\,758\,B to empirical objects}
\label{subsec:empi}

The Y3/J2, J2/J3, H2/H3, and K1/K2 flux ratios provide a clear detection of water and methane absorptions around 1.15, 1.6, and 2.3~\mic in the atmosphere of the brown-dwarf companion. We compared its 1--2.5~\mic SED to those of 101 T0--T8 field dwarfs with near-infrared spectra taken from the SpeXPrism library \citep{2014ASInC..11....7B}. The mean flux $F_{k,i}$ and error $\sigma_{F_{k,i}}$ associated to each template spectrum $k$ and filter passband $i$ was estimated and compared to the companion SED $f$ and error $\sigma_{f}$ using the $G''$ goodness-of-fit indicator defined by \cite{2010ApJ...723..850B}: 

\begin{center}
\begin{equation}
\label{eq:G2k}
G''_{k}=\sum_{i=1}^{n}w_{i}\frac{(f_{i} - C''_{k}F_{k,i})^2}{\sigma_{f_{i}}^2 + (C''_{k}\sigma_{F_{k,i}})^2}
\end{equation}
\end{center}

where $C''_{k}$ is a renormalization factor applied to the template SED $k$  which minimizes $G''_{k}$. $w_{i}$ is the renormalized FWHM $\Delta\lambda_{i}$ of each filter $i$ following:   

\begin{center}
\begin{equation}
\label{eq:eq2}
w_{i}=\frac{\Delta\lambda_{i}}{\sum_{j=1}^{n}\Delta\lambda_{j}}
\end{equation}
\end{center}
 
\begin{figure}
\begin{center}
 \includegraphics[width=\columnwidth]{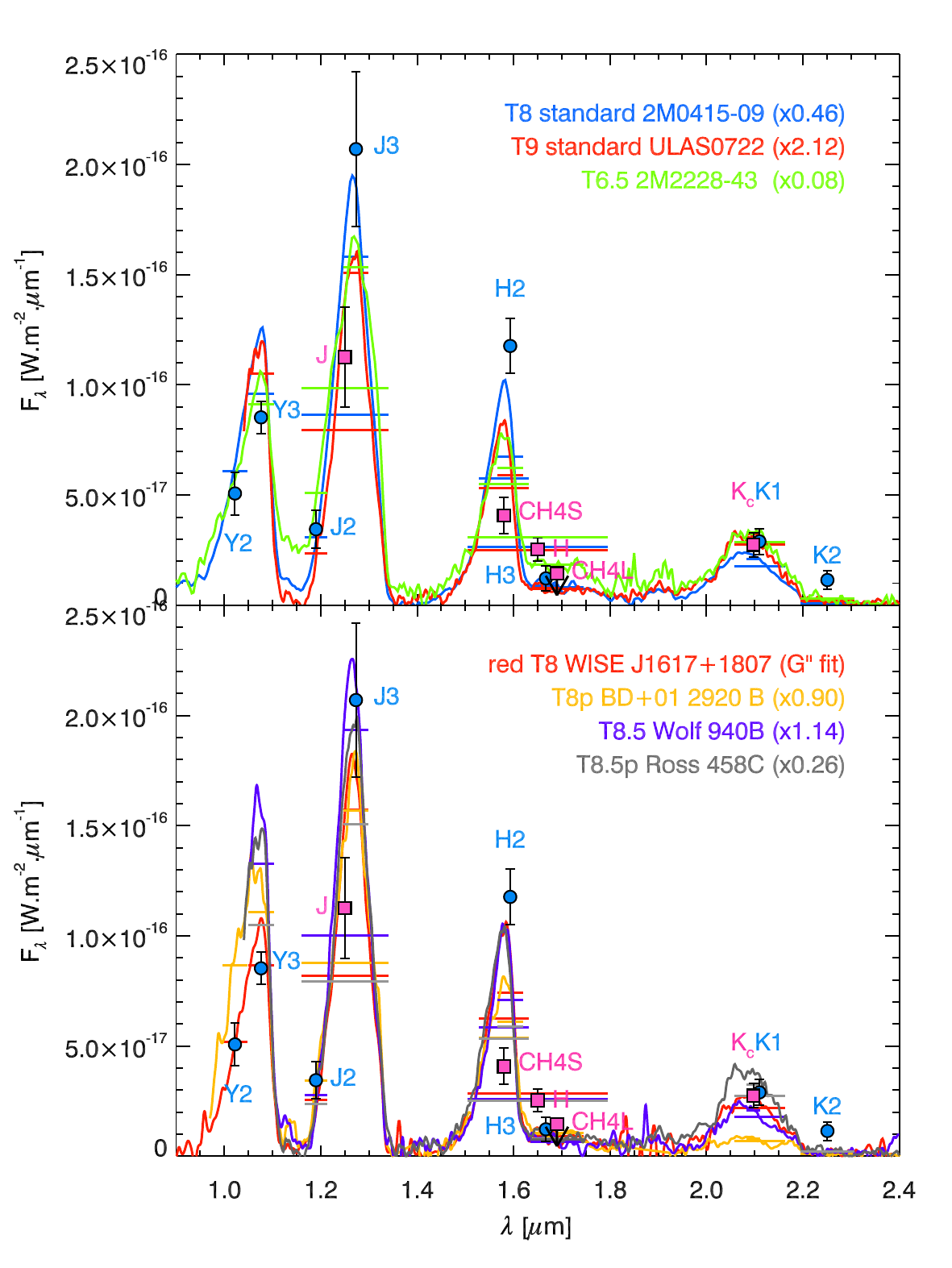}
  \caption{Comparison of the 1--2.5~\mic spectral-energy distribution of GJ\,758\,B to those of T8, T9 standard, benchmark companions, and to the red T8 dwarf WISEJ1617+1807 \citep{2011ApJ...735..116B}. The large blue circles represent our new IRDIS measurements, while the large pink squares represent the measurements from \citet{janson2011}. The horizontal lines correspond to the expected fluxes of the empirical objets in each filter bandpass.}
\label{Fig:FigSEDcompemp}
\end{center}
\end{figure}

\begin{figure}
 \centering
 \includegraphics[width=\columnwidth]{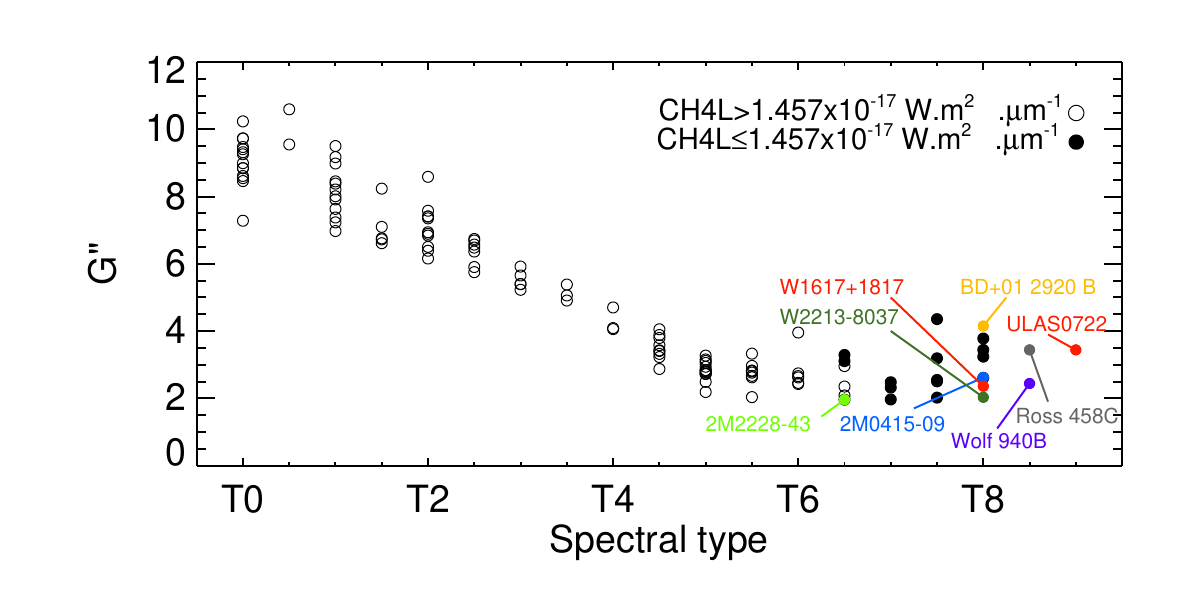}
  \caption{$G''$ values inferred from the comparison of SEDs of T dwarfs (generated from SpecXPrism spectra) with with GJ\,578\,B's. The re-normalized SEDs whose flux in the CH4L passband respect the upper limit set for GJ\,758\,B are reported as filled dots. Those who don't are reported with open circles. The G'' values for the objects considered in Fig.~\ref{Fig:FigSEDcompemp} are overlaid. We also report the value for the red T8 dwarf WISEP\,J231336.41-803701.4 whose SED, along with the one of the red T8 WISEP\,J161705.75+180714.0 provide the best visual fits to the SED of the companion.}
\label{Fig:Gprime_vs_Sptype}
\end{figure}

The indicator enables us to compare SEDs with an inhomogeneous wavelengths sampling and with measurement errors on both the templates and the object. We rejected solutions which exceeded the upper limit of the flux into the CH4L passband \citep{janson2011}. The $G''$ indicator is minimized for the T6.5 dwarf \object{2MASS J22282889-4310262} \citep{2004AJ....127.2856B} which is known to experience wavelength-dependent photometric variability \citep{2012ApJ...760L..31B}. The comparison is shown in the upper panel of Fig.~\ref{Fig:FigSEDcompemp}, and we report the $G''$ values as a function of spectral type in Fig.~\ref{Fig:Gprime_vs_Sptype}. When flux-calibrated and scaled to the distance of GJ\,758\,B \citep[using the parallax of ][]{2012ApJ...752...56F}, the spectrum of \object{2MASS J22282889-4310262} is over-luminous and a multiplication factor of 0.08 must be applied to fit the companion SED. This indicates that GJ\,758\,B is most likely later than T6.5. The variation of $G''$ with the spectral type also clearly confirms that the companion is later than T5. This is in agreement with the conclusions of \cite{janson2011}.

We show in the lower panel of Fig.~\ref{Fig:FigSEDcompemp} the spectra of standard T8 and T9 dwarfs \citep{2004AJ....127.2856B,2010MNRAS.408L..56L} with measured trigonometric parallaxes and fluxes brought to the distance of the GJ\,758 system. The companion SED is midway between the renormalized SED of the T8 and T9 standards. Nevertheless, the templates fail to reproduce the J3, H2, and K2 fluxes simultaneously. The companion also appears to have a luminosity intermediate between these two objects. Its $J$ and $H$ band absolute magnitudes agree well with the mean values reported in \cite{2013Sci...341.1492D} for T8-T8.5 objects.

The causes of the peculiar SED of GJ\,758\,B are unclear. The companion spectrophotometric properties could be related to a non-solar composition, or a surface gravity different than those of the standard T8-T9 dwarfs. Both parameters produce opposite effects on 1--5~\mic SEDs that are difficult to disentangle \citep[e.g.][]{2010ApJ...710.1627L}. We used the spectra of wide companions to stars with known age, and metallicity to investigate the effect of peculiar atmospheric parameters, making the assumption that these objects share the same composition as their host star. 

\object{Ross\,458}\,C \citep{2010MNRAS.405.1140G, 2010A&A...515A..92S} appears as the only object with an estimated age (150--800~Myr) younger than the typical field dwarf ages ($\gg$500 Myr) which has an estimated \Teff \citep[625-755 K][]{2010ApJ...725.1405B, 2011MNRAS.414.3590B} and near-infrared spectral type (T8.5p) in the same range as that of GJ\,758\,B \citep[][]{janson2011}. It is also reported to have a super-solar metallicity \citep[Fe/H = +0.2--0.3;][]{2010ApJ...725.1405B}, e.g. similar to that of GJ\,758\,A (+0.2 dex, see Sect.~\ref{sec:stellar_parameters}). The spectra of both objects are also compared in the lower panel of Fig.~\ref{Fig:FigSEDcompemp}. The spectrum of Ross\,458\,C from \citet{2011MNRAS.414.3590B} represents the SED of GJ\,758\,B less well than the T8 standard. Its enhanced flux at $K$-band suggests that the two companions do not span the  same surface gravity and/or metallicity interval. 

We considered the opposite case of the peculiar T8 companion to the metal-poor (${\rm [Fe/H]} = -0.38 \pm 0.06$~dex) G-type star \object{BD+01\,292} \citep{2012MNRAS.422.1922P} and of the T8 companion to the sdM1.5+WD binary Wolf\,1130 (${\rm [Fe/H]} = -0.64 \pm 0.17$). Both companions have a suppressed flux at $K$-band, possibly due to the enhanced collision-induced absorption of H$_{2}$ encountered into clear/low-metallicity/higher-pressure atmospheres \citep{1994ApJ...424..333S, 1997A&A...324..185B}. They clearly produce a worse fit to the SED of GJ\,758\,B than the T8 standard does. In summary, we see an opposite trend for GJ\,758\,B departure from the SED of the standard T8.

The T8.5 companion to the old (3.5--6~Gyr) solar-metallicity star \object{Wolf\,940} \citep[${\rm [Fe/H]} = -0.06 \pm 0.20$]{2009MNRAS.395.1237B} represents the $J$ band flux better, at the price of a degradation of the fit in the $Y$ band. We do not find a good fit with earlier type companions such as \object{GJ\,229}\,B (T7pec) or \object{Gl\,570}\,D (T7.5) \citep{1996ApJ...467L.101G, 2001ApJ...556..373G} and primaries with roughly solar-metallicities \citep{2014A&A...568A.121N}. 

We extended the comparison to additional peculiar dwarfs with red near-IR colors but no a priori knowledge of their age and metallicity \citep[e.g.][and references therein]{2013ApJ...777...36M}. We find that the red T8 dwarfs \object{WISEP J161705.75+180714.0} and \object{WISEP J231336.41-803701.4} \citep{2011ApJ...735..116B} provide the best fit among all other aforementioned objects. They notably represent the $Y$ band flux well compared to the other objects. \cite{2011ApJ...735..116B} note that the spectral properties of these two objects suggest cool ($\Teff = 600$~K), low surface gravity ($\logg = 4.0$), and cloudy atmospheres.

In summary, we cannot find an empirical object with known metallicity and distance that accurately represents all the near-IR narrow-band and broad-band fluxes of GJ\,758\,B simultaneously. We estimate a T8 spectral type from this comparison. The analysis is however certainly limited by the small amount of spectra of T8--T9 dwarfs with robust constraints on their age and metallicity. 

\subsection{Comparison of GJ\,758\,B to atmospheric models}
\label{subsec:atmomodels}

We compared the SED of GJ\,758\,B to four sets of atmospheric models -- BT-Settl, Exo-REM, Morley+12 and Saumon+12 -- in order to refine the estimate of \logg, \Teff, and Fe/H, and to understand its peculiar photometry. The models are described in \citet{2013MSAIS..24..128A}, \citet{2015arXiv150404876B}, \citet{2012ApJ...756..172M}, and \citet{2012ApJ...750...74S} respectively. The specificities and parameter space of the models are decribed in more details in Appendix~\ref{sec:description_atmospheric_models} and Table~\ref{Tab:atmomodchar}. We expect that the use of these different classes of models allows the best possible approach for the accurate modeling of the atmospheric parameters.

In order to account for the inhomogeneous sampling of the real SED during the fitting process, we decided to use the goodness-of-fit $G_{k}$ indicator defined by \cite{2008ApJ...678.1372C}. This indicator contains a dilution factor, $C_{k}$, similar to the $C''_{k}$ factor defined in Eq.~\ref{eq:G2k}. $C_{k}$ usually equals $(R/d)^{2}$, where $d$ is the distance of the source and $R$ its radius. Given the \emph{Hipparcos} distance of GJ\,758\,A, we were able to retrieve the optimal average object radii for each given model. 

\begin{figure}
  \centering
  \includegraphics[width=\columnwidth]{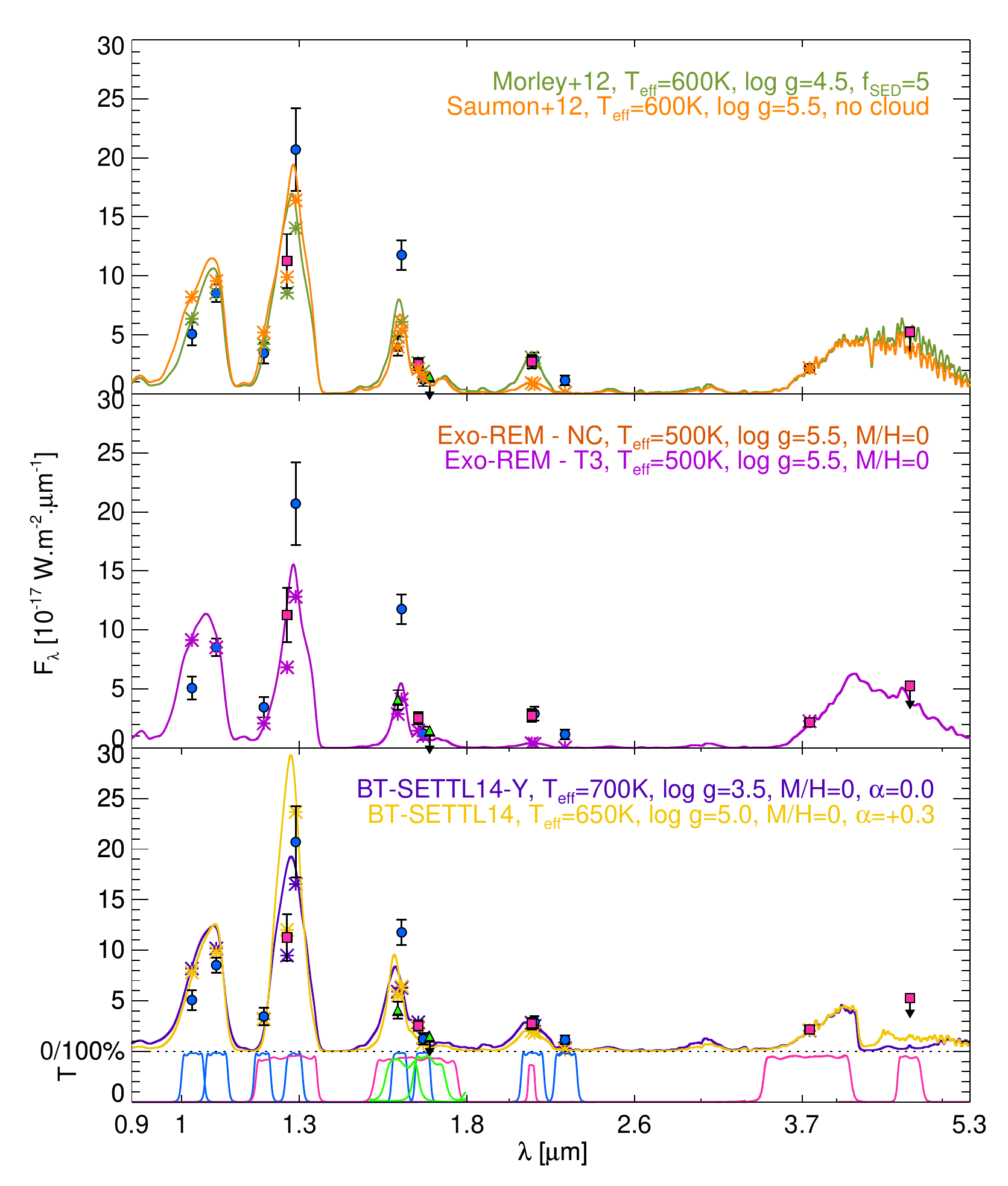}
  \caption{Comparison of the 1-5 $\mu$m spectral-energy distribution of GJ\,758\,B to the best fitting synthetic spectra from the BT-SETTL14, Morley+12, Saumon+12, and Exo-REM grids. The asterisks represent the fluxes in each bandpass expected from the atmospheric models. The Exo-REM and Exo-REM~-~NC models completely overlap because at this combination of \Teff and \logg, the cloud condensation occurs below the considered pressure grid, effectively making both models cloud-free.}
  \label{Fig:FigSEDcompsynth}
\end{figure}

\begin{figure}
  \centering
  \includegraphics[width=\columnwidth]{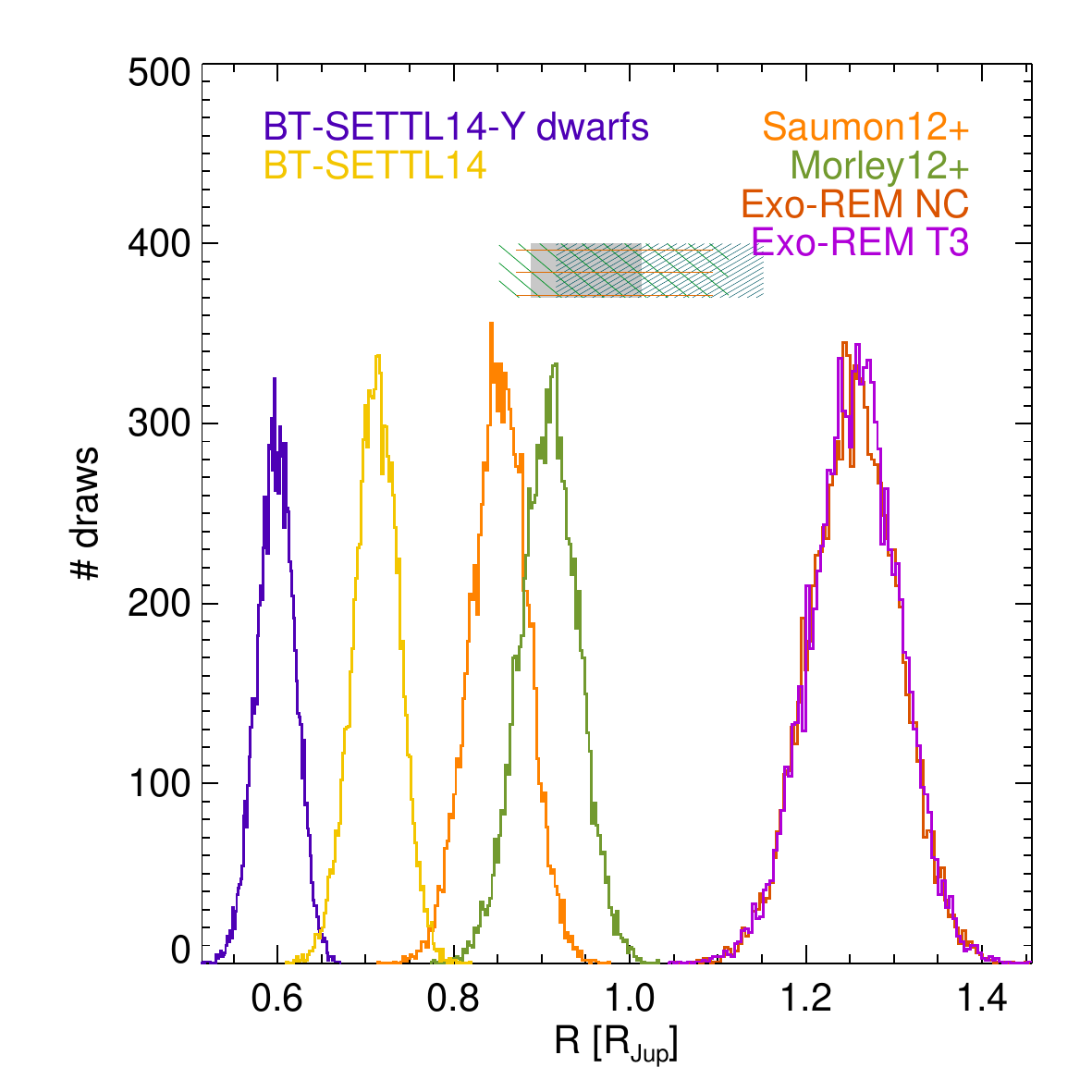}
  \caption{Histogram of radii (dilution factors $C_{k}$, directly related to radius because the distance is known) derived from the comparison of the most frequent best-fitting solution for each Monte-Carlo simulation of the SED of GJ\,758\,B. The hatched areas correspond to the range of radii predicted for the estimated \Teff and age of the system by the \citet{2008ApJ...689.1327S} models with cloudy (blue hatches), hybrid clouds (red hatches), and cloudless (green hatches; covering $[{\rm M/H}]$ of 0, 0.3, and -0.3~dex) atmospheres considered as boundary conditions. The shaded zone correspond to the predictions of the COND models \citep{baraffe2003}.}
  \label{Fig:FigSEDradius}
\end{figure}

Confidence levels cannot be derived directly from $G$. Therefore, we followed the approach of \cite{2008ApJ...678.1372C} to determine the most meaningful fitting solution of each model grid. For each photometric data point of the object, we generated a normal Monte-Carlo (MC) distribution of 10\,000 draws with mean values of $f_{i}$ and standard deviations of $\sigma_{i}$. The $G_{k}$ values were computed for each of the resulting 10\,000 SEDs. We computed for each models of the grid, the fraction of the 10\,000 MC simulated SEDs that were best fitted by this given model. This $f_{MC}$ indicator, ranging from 0 to 1, enables to test the significance of any fitting solution. The models with the highest $f_{MC}$ value represents the most significant solution. However, we note that $f_{MC}$ is sensitive to the sampling and extent of the model grid. Therefore, despite the criterion is usefull to estimate the robustness of a given solution within a grid, it should not be used to evaluate the quality of the solutions found with different grids. We performed a visual inspection of the three solutions with the highest $f_{MC}$ for each model grid, but only reported the atmospheric parameters and $f_{MC}$ of the most probable solution in Table \ref{Table:SEDfitresults}.

The Monte-Carlo method works as long as the errors associated to $f_{i}$ are uncorrelated. In the case of GJ\,758\,B, the errors associated to the flux-calibrated SED of the object combine uncorrelated errors corresponding to the  companion contrast values associated to each filter to a correlated error arising for the flux-calibrated spectrum of the star. We accounted for both sources of errors in our MC simulations by multiplying the 10\,000 MC SEDs of the companion by $10^{-0.4 \times \textsl{N}(\mu=0,\sigma_{ph})}$ with $\textsl{N}$ an additional MC normal distribution of 10\,000 values with mean values of 0 and standard deviation $\sigma_{ph}$ equal to the magnitude error on the flux scaling of the companion spectrum. We took $\sigma_{ph} = 0.03$~mag, which corresponds to the highest photometric error on the SED of the star.

The results of the fits are reported into Table~\ref{Table:SEDfitresults} and shown in Fig.~\ref{Fig:FigSEDcompsynth}. The solutions with the highest $f_{MC}$ always correspond to the solution with the minimum $G$. The corresponding dilution factors inferred from our MC simulations for the most probable fitting solution (highest $f_{MC}$) are shown in Fig.~\ref{Fig:FigSEDradius}. No one model represents well the whole SED, especially the J3 and H2 fluxes. The Morley+12 models provide the best fits according to the $G$ indicator. The three most significant solutions ($>$65\% of the solutions) found with these models correspond to log g=4.0-4.5 and $\Teff = 550-600$~K. The cloud-free Saumon+12 models only provides a better fit to the $J$ band flux. But their poorer representation of the other bands indicate that clouds are still needed in the photosphere to reproduce the GJ\,758\,B SED. Visually, the BT-SETTL14 models seem to provide a better fit to the $H$ band flux. The flux drop at $M_{s}$ band in the BT-SETTL14 models is in better agreement with the upper limit found by \cite{janson2011}. New deeper observations at $M$ band of the GJ\,758 system could help to further discriminate the models. The BT-SETTL14 models do not provide any meaningful constraints on the log g. A re-analysis with a classical $\chi^{2}$ confirms the conclusions. The radii (dilution factors) needed to adjust the surface flux predicted by the models onto the apparent flux of the companion are unphysical in the case of the BT-SETTL14 models. This may indicate that the \Teff of GJ\,758 could be lower than the one corresponding to the best fit. The Exo-REM models fail to represent correctly the SED of the companion, especially in the $Y$, $H$, and $K$-bands.

We conclude that the companion has $\Teff = 600 \pm 100$~K from the above analysis. This is in good agreement with \cite{janson2011} whose analysis relied on the models of \cite{2006ApJ...640.1063B} extended to colder temperatures (Hubeny \& Burrows, in prep.). The Morley+12 parametric model points toward a low surface gravity, in agreement with the hints found in Sect.~\ref{subsec:empi}. But the inexistent exploration of the effect of the metal-enrichment in these grids of models associated to model uncertainties certainly biases the analysis. A low-resolution spectrum of the source would be needed to determine with good confidence \logg and M/H.

From the \Teff and the derived age for the system ($3^{+3}_{-2}$~Gyr, see Sect.~\ref{sec:stellar_parameters}), we estimate a mass of $23^{+17}_{-13}$~\MJup for GJ\,758\,B using the BT-SETTL13 grid of models \citep{2013MSAIS..24..128A}. This value is in the low range of the masses inferred by \citet{janson2011}, as a direct consequence of our estimated age range for the system that is slightly younger than the one they considered.

Finally, we report in Table \ref{Tab:evolm} the predictions from the \cite{baraffe2003} and \cite{2008ApJ...689.1327S} models. The models predict radii corresponding to the estimate \Teff in the range 0.80--1.21~$R_{Jup}$ which are marginally consistent with the radii derived from the SED fit with the Exo-REM, Saumon12+ and Morley12+ models. They are still 25 and 11\% larger than those infered from the SED fit with the BT-SETTL14-Y dwarfs and BT-SETTL14 models respectively (see Fig.~\ref{Fig:FigSEDradius}). The difference may arise from the different boundary conditions considered for the evolutionary models and the atmospheric models used for the SED fit. Nevertheless, it is more likely that the \Teff derived from the SED fit is slightly overestimated by the SETTL models (by 100--200 K) and lead to this inconsistency.

\begin{table*}[t]
\centering
\caption{\label{Table:SEDfitresults} Fitting solutions with the highest $f_{MC}$ values for the GJ\,758\,B SED and the three sets of atmospheric models using the $G$ goodness-of-fit indicator.}
\begin{tabular}{lllllllll}
  \hline
  Model        & \Teff&\logg&[M/H]& $[\alpha]$ & $f_{SED}$ & R & G & $f_{MC}$ \\
  \hline
  BT-SETTL14-Y & 700 & 3.5 & 0.0 & 0.0 & n/a & 0.60 & 1.80 & 0.55 \\
  BT-SETTL14   & 650 & 5.0 & 0.0 & 0.3 & n/a & 0.71 & 1.60 & 0.20 \\
  Exo-REM - NC & 500 & 5.5 & 0.0 & 0.0 & n/a & 1.26 & 4.83 & 0.85 \\
  Exo-REM - T3 & 500 & 5.5 & 0.0 & 0.0 & n/a & 1.26 & 4.83 & 0.63 \\
  Morley+12    & 600 & 4.5 & 0.0 & 0.0 & 5.0 & 0.91 & 1.39 & 0.30 \\
  Saumon+12    & 600 & 5.5 & 0.0 & 0.0 & n/a & 0.85 & 2.70 & 0.74 \\
  \hline
\end{tabular}
\end{table*}

\begin{table}[t]
  \centering
  \caption{\label{Tab:evolm} \Teff and radius predictions from the \cite{baraffe2003} and \cite{2008ApJ...689.1327S} models}
  \begin{tabular}{ccccc}
  \hline
  Models     & Boundary  & M/H   & $R$                 & Mass    \\
             &           & (dex) & (\RJup)             & (\MJup) \\
  \hline
  Saumon+08  & No cloud  & 0     & $0.96^{0.14}_{0.10}$  & $24^{+16}_{-13}$ \\
  Saumon+08  & No cloud  & $+$0.3& $0.97^{0.15}_{0.10}$  & $24^{+15}_{-14}$ \\
  Saumon+08  & No cloud  & $-$0.3& $0.95^{0.15}_{0.10}$  & $24^{+16}_{-13}$ \\
  Saumon+08  & Cloudy    & 0.0   & $1.01^{0.15}_{0.10}$  & $21^{+14}_{-11}$ \\
  Saumon+08  & Hybrid    & 0.0   & $0.96^{0.13}_{0.10}$  & $24^{+14}_{-13}$ \\
  COND       & AMES-COND & 0.0   & $0.97^{0.05}_{0.08}$  & $21^{+11}_{-10}$ \\
  \hline
  \end{tabular}
\end{table}

\subsection{Nature of the new candidate companion}
\label{sec:nature_new_candidate}

\begin{figure}
  \centering
  \includegraphics[width=\columnwidth]{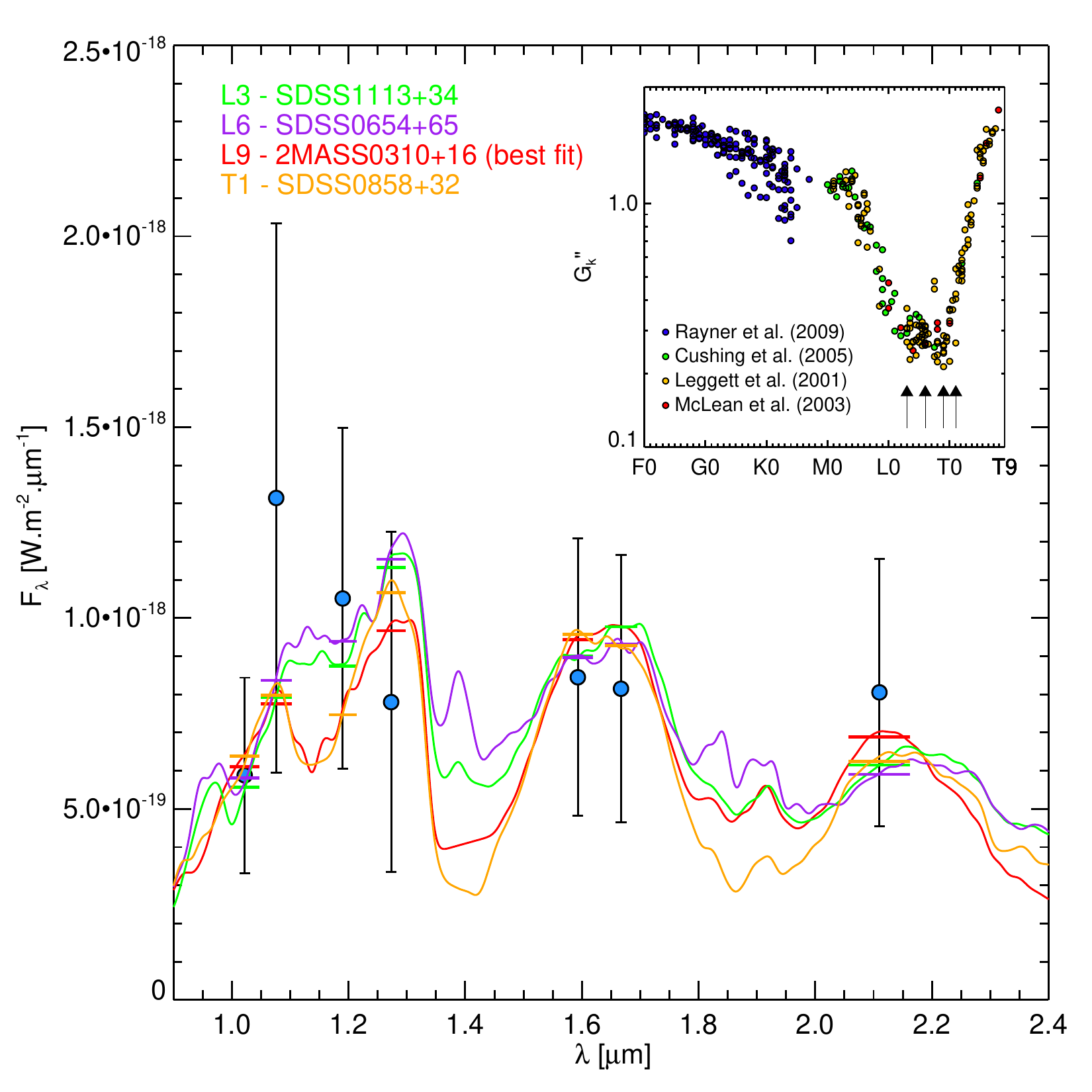}
  \caption{Comparison of the flux of the newly identified candidate (blue circles) with SEDs of different sub-stellar objects of spectral types L3, L6, L9 (best fit) and T1. For each spectral type, we plot the object that provides the best fit according to the $G''_{k}$ indicator. The inset plot at the top shows the $G''_{k}$ values as a function of spectral type for $\sim$400 objects taken from various libraries \citep{leggett2001,mclean2003,cushing2005,rayner2009}. The vertical arrows indicate the spectral type of the plotted SEDs.}
  \label{fig:new_candidate_photometry}
\end{figure}

The detection of a new candidate companion around a star with an already known companion is particularly interesting. It is going to become very common with the new generation of high-contrast imagers because of the boost in sensitivity that they provide at smaller angular separations. For the new candidate detected in our IRDIS data, we make use of the large multi-wavelength coverage ($Y$- to $K$-band) to perform a photometric analysis. 

Although the error bars on the photometry of the new candidate are large (>0.5~mag, see Table~\ref{Tab:MagAb}), we attempt a first-order estimation of its spectral type by comparing its observed flux with the SEDs of stellar and sub-stellar objects. The star data are taken from the IRTF stellar library\footnote{\url{http://irtfweb.ifa.hawaii.edu/~spex/IRTF_Spectral_Library/}} \citep{cushing2005,rayner2009}, while the brown dwarf data are taken from the NIRSPEC brown dwarf spectroscopic survey \citep{mclean2003} and from \citet{leggett2001}. For the comparison, we use the $G''_{k}$ indicator as in Sect.~\ref{subsec:empi}. The results are presented in Fig.~\ref{fig:new_candidate_photometry}, where we show the $G''_{k}$ values as a function of spectral types (inset), and the SEDs of four objects that can fit equally well the photometry of the new candidate within the error bars.

The $G''_{k}$ distribution shows a rather flat minimum in the L3--T1 range, indicating that our candidate could likely be of sub-stellar nature. However, reaching a final conclusion is difficult from our current data because of the significant uncertainties on the photometry. As showed in Fig.~\ref{fig:new_candidate_photometry}, the candidate photometry is compatible with mid-L to early-T types, but late-M and early-L (not showed) would also provide decent fit. We note that the low galactic latitude of GJ\,758 (+8~deg) increases significantly the probability of background contamination, particularly with late M stars, which are the main source of contamination at high-contrast \citep[e.g.][]{chauvin2015}.

Other possibilities for the nature of the candidate could include Solar System bodies, such as asteroid and trans-neptunian objects, or extra-galactic objects. However, an asteroid basically reflects the near-IR light from the Sun, resulting in a very flat G2V spectrum that is not compatible with the photometry. In addition, these objects would be characterized by a very large proper motion of several mas to several dozens of mas per second. Our observations taken over two consecutive nights completely rule out this possibility. On the other hand, extra-galactic sources such as galaxies are another possibility, but they would be resolved even at significant redshifts by the very fine plate scale of IRDIS ($\sim$12.25~mas, see Table~\ref{tab:astrometric_calibration}). The point-like structure of the candidate also rules out this possibility.

In conclusion, we cannot rule out the possibility that the new companion is indeed bound to GJ\,758, since its photometry is broadly compatible with L-type objects. A second epoch will be required to clear any possible doubt. Given the high proper motion of the star ($\sim$180~mas/yr), a confirmation of the status of this candidate is already possible.

\section{Astrometry and orbital properties}
\label{sec:astrometry_orbital_properties}

\subsection{Least-Square Monte Carlo orbital fitting}
\label{sec:lsmc_orbital_fitting}

\begin{figure*}
  \subfloat[][]{
  \includegraphics[width=0.32\textwidth]{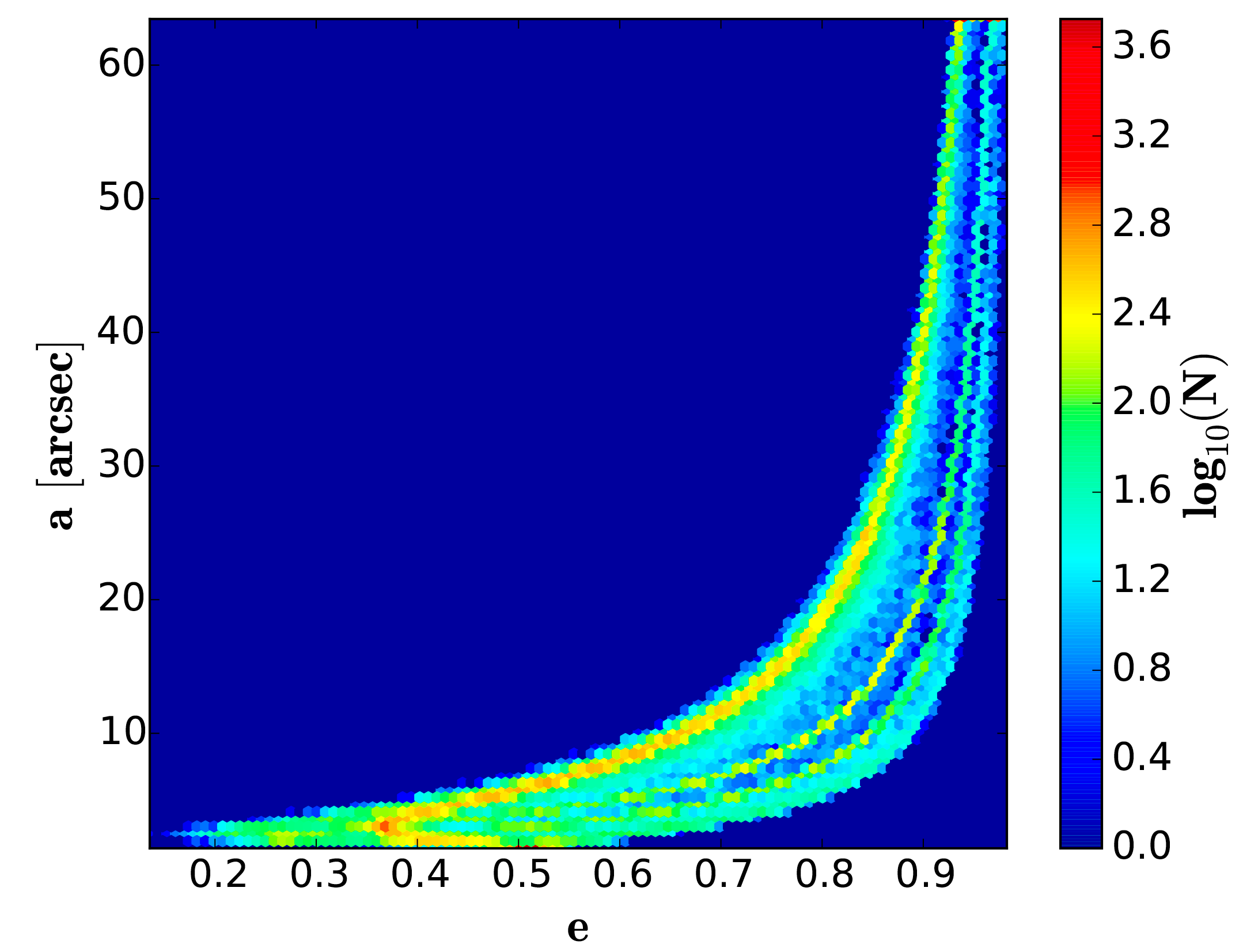}
  \label{fig:a-vs-e}
  }
  \subfloat[][]{
  \includegraphics[width=0.32\textwidth]{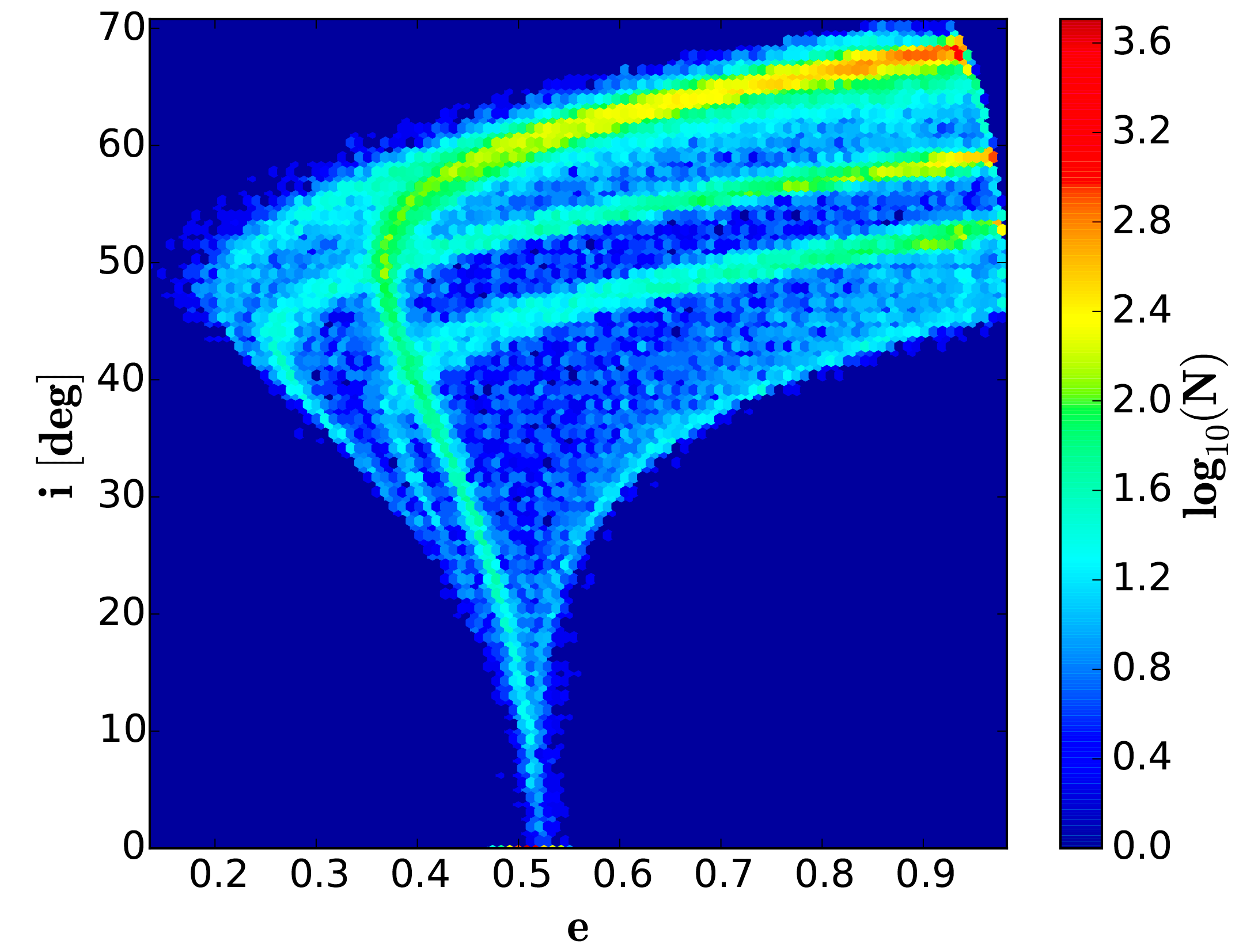}
  \label{fig:i-vs-e}
  }
  \subfloat[][]{
  \includegraphics[width=0.32\textwidth]{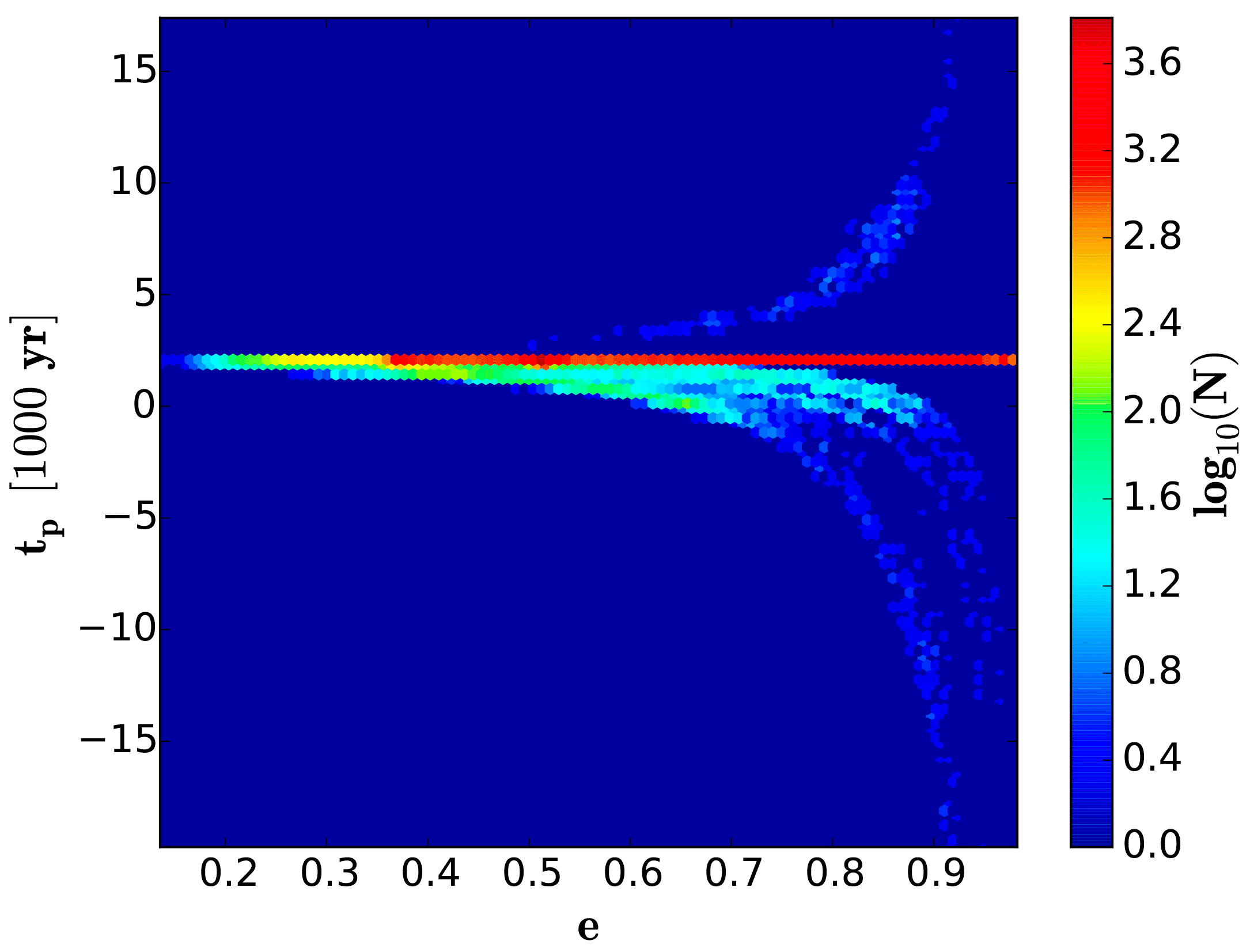}
  \label{fig:T0-vs-e}
  }
  \caption{Semi-major axis, inclination and time of periastron passage as function of eccentricity for all solutions with $\chi^2_{red} \ \leq \ 2$ out of 5\,000\,000 runs of our LSMC fit. Logarithmic density of solutions is indicated by color.} 
\label{fig:orbit-corr}
\end{figure*}

\begin{figure}
  \centering
  \includegraphics[scale=0.51]{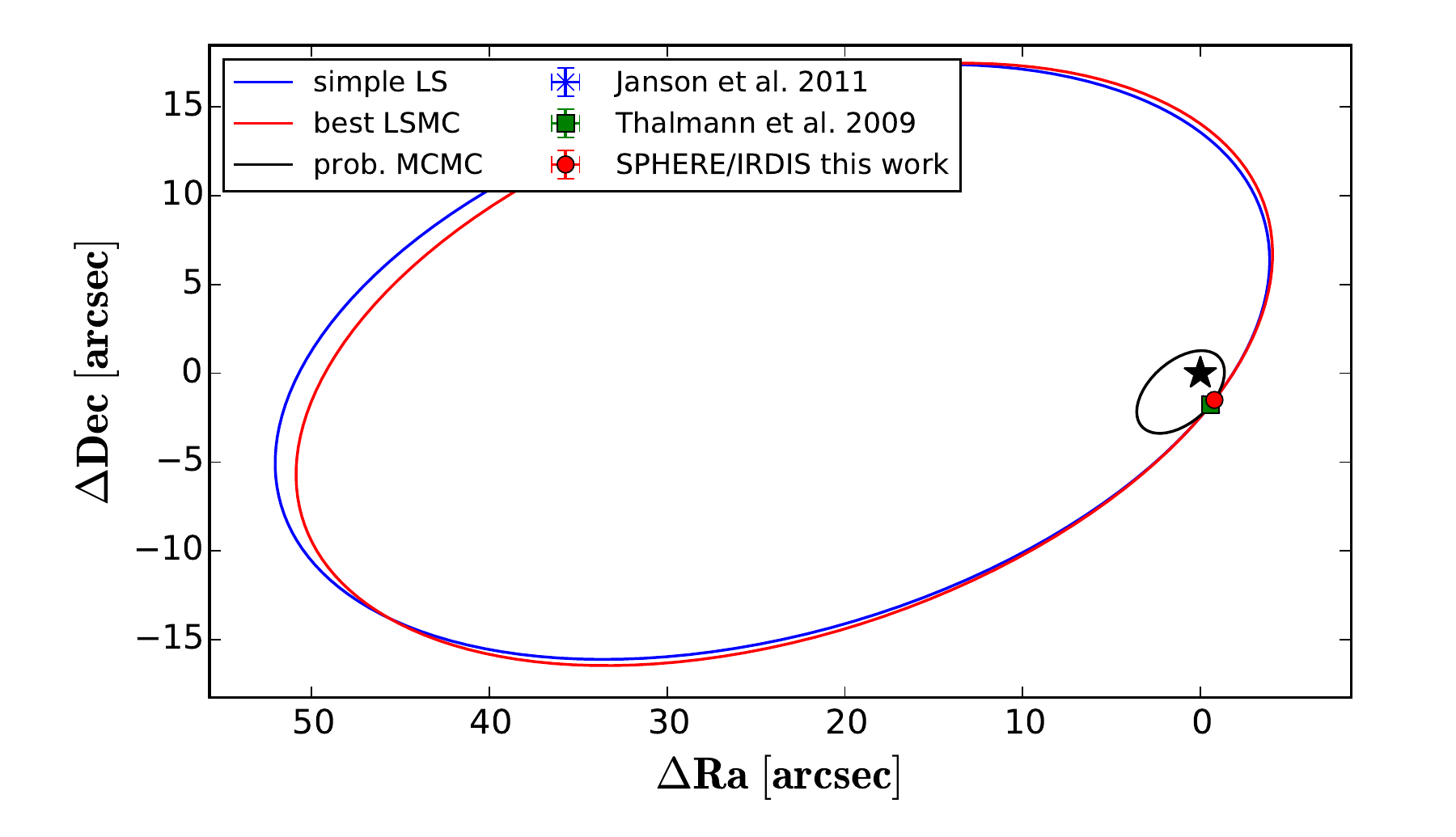}
  \includegraphics[scale=0.51]{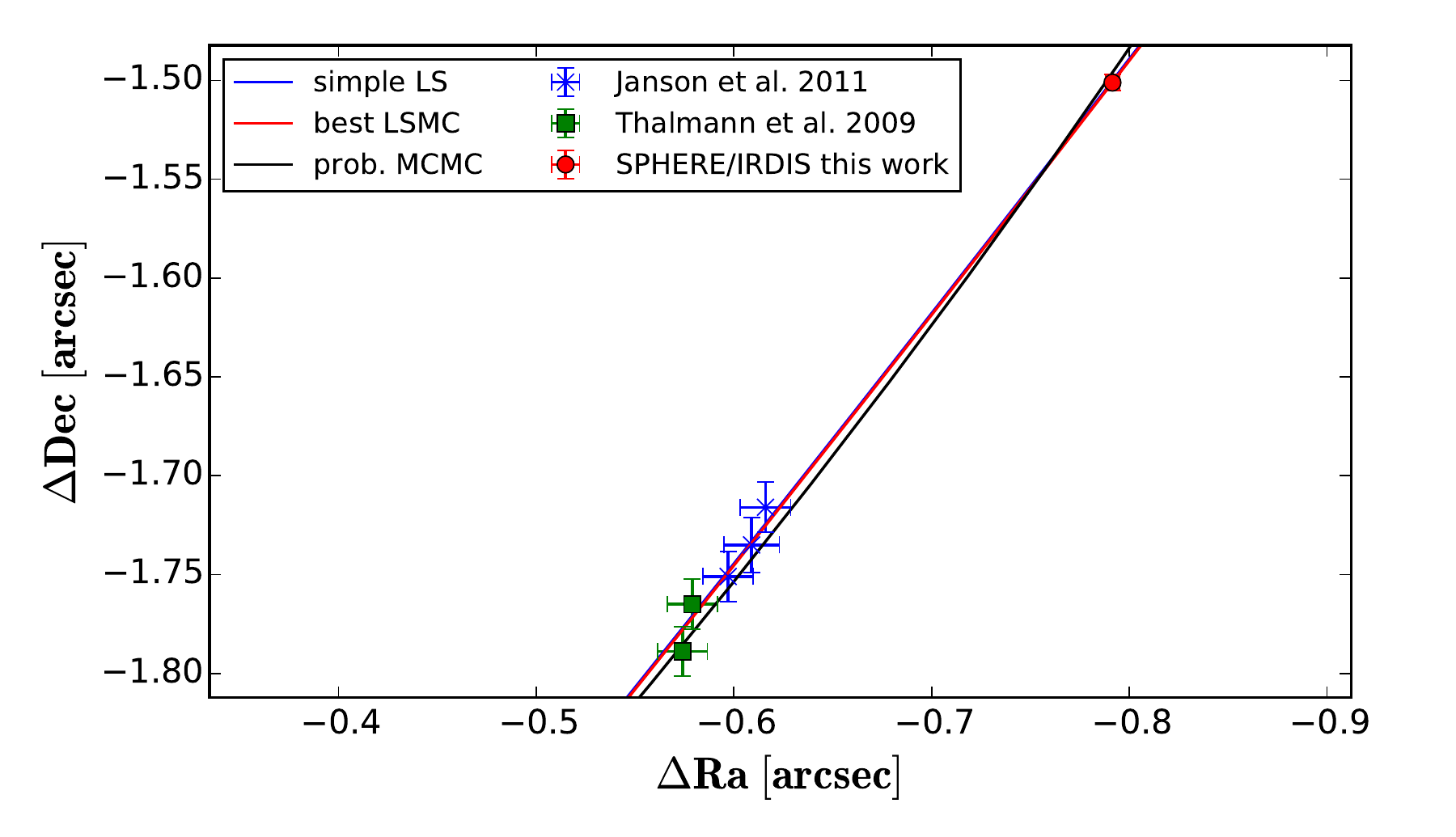}
  \includegraphics[scale=0.46]{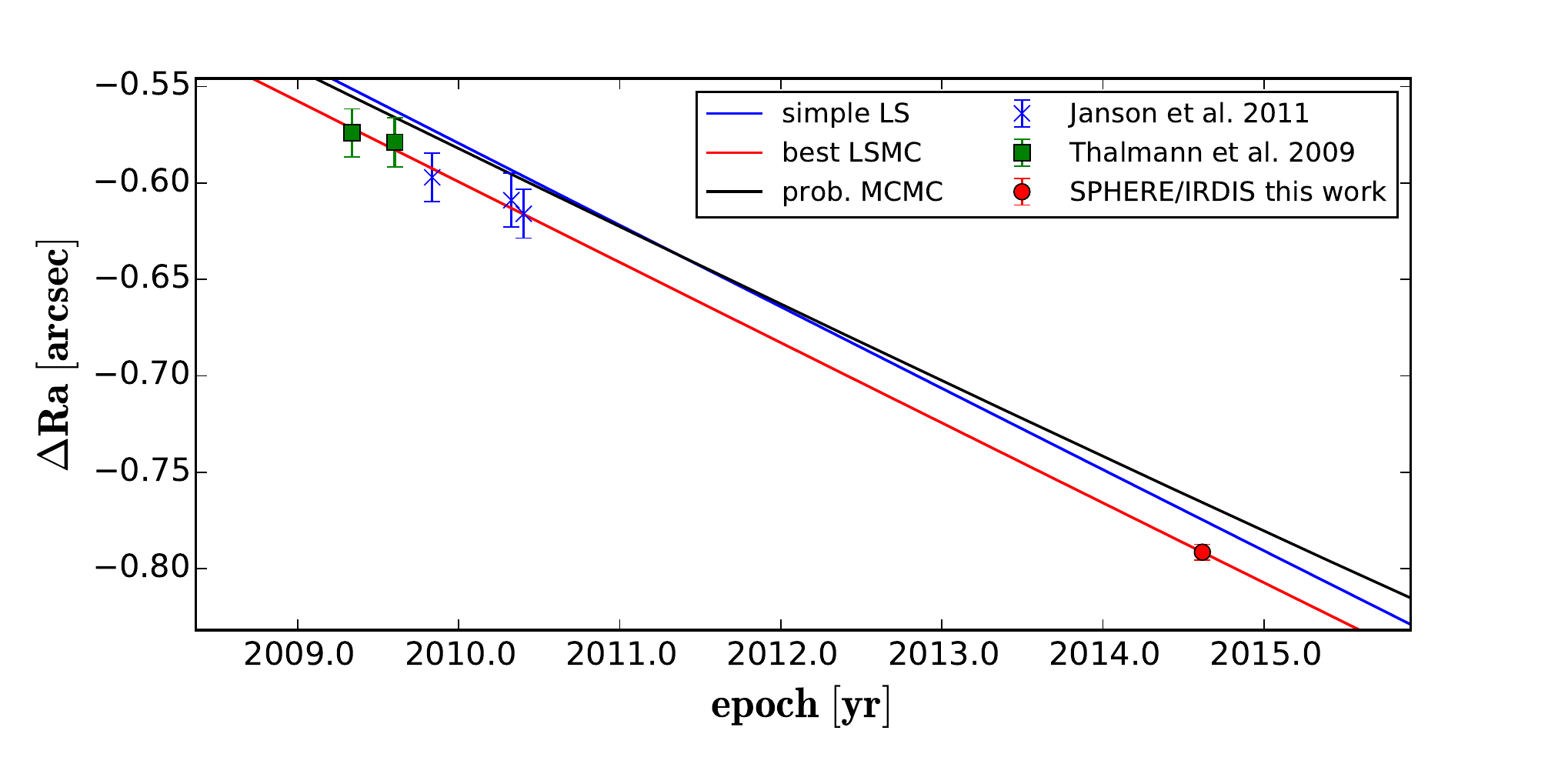}
  \includegraphics[scale=0.46]{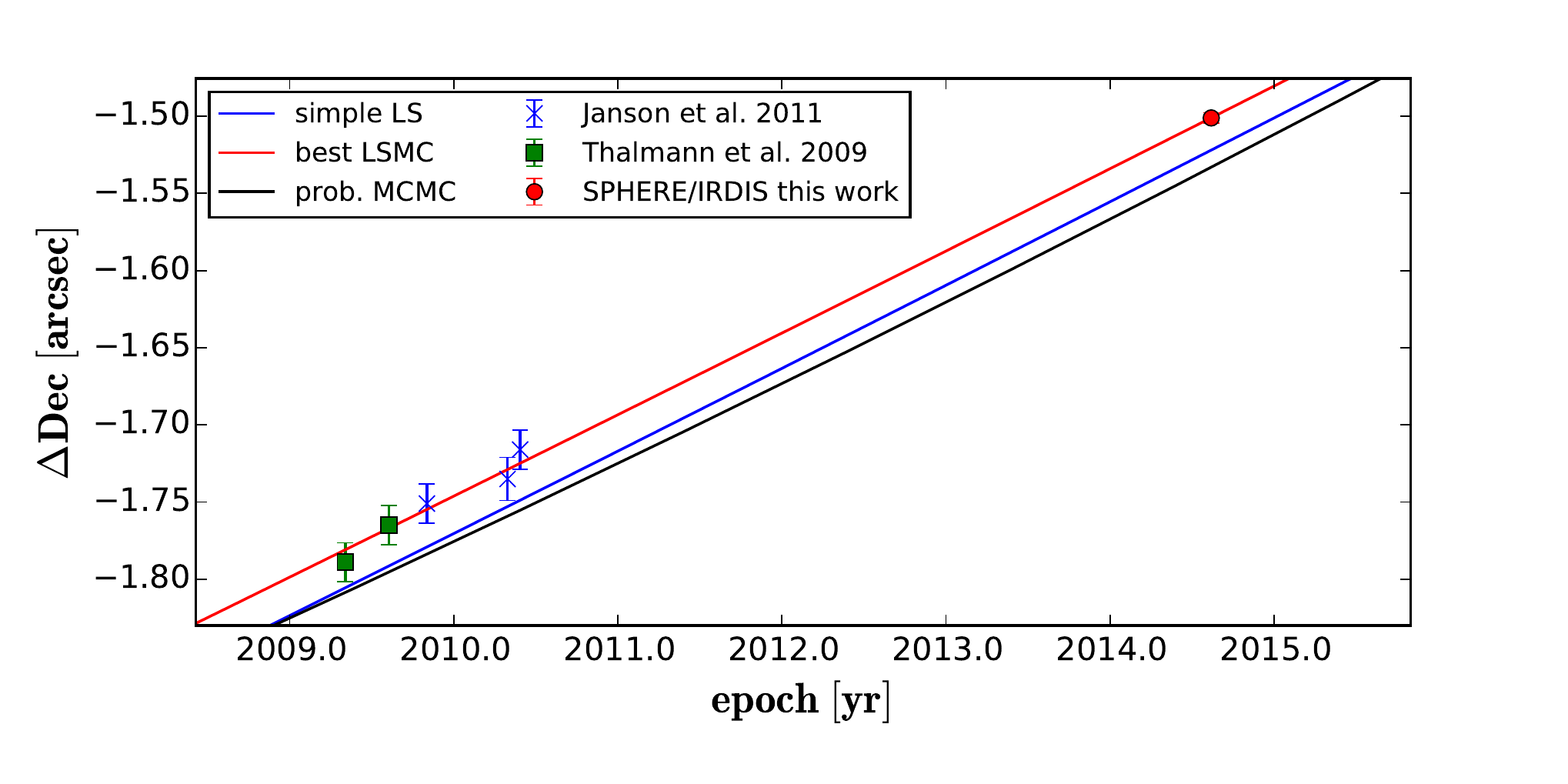}
  \caption{Best fitting orbits recovered with simple least squares fitting as well as LSMC fitting. In addition, we show a probable orbit with orbital elements close to the peak values recovered by our MCMC fit (see Sect.~\ref{sec:mcmc_orbital_fitting}). Solid lines represent the apparent orbits. The corresponding orbital elements are listed in Table~\ref{gj758_stats}. We show the data points taken with Subaru/HiCIAO (green squares) as given in \citet{thalmann2009}, as well as the data points taken with Subaru/HiCIAO, Gemini/NIRI and Keck/NIRC2 (blue crosses) given in \citet{janson2011} together with our SPHERE/IRDIS measurement (red circle).} 
  \label{fig:orbits}
\end{figure}

\begin{figure}
  \includegraphics[width=1.0\columnwidth]{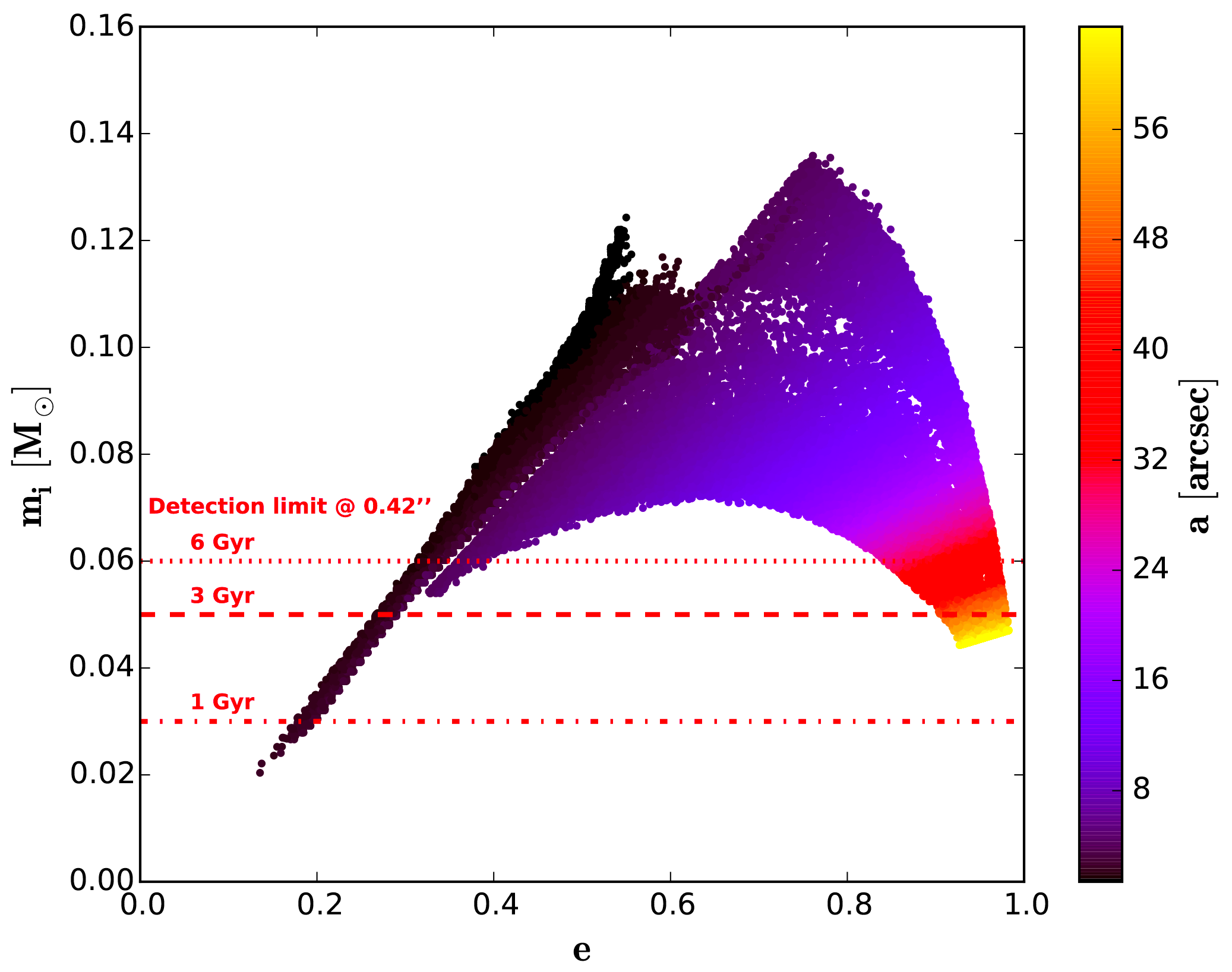}
  \caption{Minimum mass of an unseen inner companion that would cause a false positive eccentricity signal in the relative astrometry of GJ\,758\,A and B by astrometric displacement of GJ\,758\,A due to their common orbit around the center of mass of the system. The minimum mass is a function of the eccentricity and semi-major axis of the A/B system as well as the maximum epoch difference of all astrometric measurements. Shown are such minimum masses for all orbits with $\chi^2_{red} \ \leq \ 2$ which were recovered for the A/B system. Using our deep SPHERE/IRDIS observations as well as the AMES-COND models \citet{baraffe2003} we also show the detectable minimum mass at the angular separation at which such a putative inner companion would need to reside. We can exclude the presence of an object that would introduce a false positive eccentricity in all cases.} 
  \label{fig:pearce-gj758}
\end{figure}

We used the new IRDIS astrometric measurement to put constraints on the orbital solution of the system. In previous studies by \citet{thalmann2009} and \citet{janson2011}, it was already shown that the system presents significant orbital motion and Monte-Carlo simulations were used to get a first estimate of the orbital elements. In this study we used first a Least-Squares Monte-Carlo (LSMC) approach to study the parameter space of possible orbits. For this purpose we created $5 \times 10^6$ sets of orbital elements, which were drawn from uniform distributions. These sets of orbital elements were then used as starting points for a least-squares minimization routine. The method is described in detail in \citet{ginski2013}. To limit the parameter space we fixed the total mass of the system to the nominal value of $\sim$1~\MSun: 0.97~\MSun for the star \citep{takeda2007} and $\sim$0.03~\MSun for the companion at the probable age of the system. In addition, we limited the semi-major axis to values smaller than 63.45\arcsec (1000~AU at a distance of 15.76~pc). This is assuming that the system is long-term stable against disruption in the galactic disk as described in \citet{close2003}. Given the high age of the system ($3^{+3}_{-2}$~Gyr, see Sect.~\ref{sec:stellar_parameters}) and the fact that we still find the companion close to the host star, this assumption seems reasonable.

The results of our simulations are shown in Fig.~\ref{fig:orbit-corr}. We do not show the results for the longitude of the ascending node and the argument of the periastron, since they are not well constrained yet by the available astrometry. In Fig.~\ref{fig:orbits} we show the best fitting orbit solution that was recovered by the LSMC orbit fit. The corresponding orbital elements are shown in Table~\ref{gj758_stats} alongside with the results recovered by our Markov-Chain Monte-Carlo simulation discussed in the following section.

Since the orbit does not show significant curvature yet, we cannot put an upper limit to the semi-major axis or the eccentricity. However, we find a lower limit of 0.14 for the eccentricity and 21.9~AU (1.39\arcsec) for the semi-major axis. In general, the semi-major axis of possible orbits scales with the eccentricity as can be seen in Fig.~\ref{fig:a-vs-e}. The minimum values of the semi-major axis and the eccentricity, as well as the general behavior of the well fitting orbit solutions, is consistent with the results presented in \citet{janson2011}, which were derived from simple Monte-Carlo simulations.

In Fig.~\ref{fig:i-vs-e} we show the inclination of possible orbital solutions as function of eccentricity. For close to face-on orbits (inclination close to 0~deg) we can constrain the eccentricity of the orbit to values between 0.47 and 0.55. This range becomes continuously larger with increasing orbit inclination. For an inclination of $\sim$50~deg, the full range of recovered eccentricities gives results consistent with the astrometric measurements. We can put an upper limit on the inclination of 70.8~deg, i.e. we can exclude edge-on orbit solutions. If we compute a simple median of the recovered orbit inclinations we get a value of $58.9 \pm 18.8$~deg. This is, within the given uncertainties, consistent with the interval found in \citet{janson2011}. Inclinations smaller than 40~deg correspond to small semi-major axes, with an upper limit of 77.5~AU (4.92\arcsec), while for larger values of the inclination, orbit solutions with the full range of recovered semi-major axes are possible.

Finally, in Fig.~\ref{fig:T0-vs-e} we show the times of the periastron passage that we recovered from our simulations. The vast majority (87.5\%) of our solutions passes the periastron between the years 2000 and 2065. The solutions that show the periastron passage at the time of the observations are generally highly eccentric and have large semi-major axes, which would explain that no curvature of the orbit is observed yet. 

While these solutions fit the orbit very well geometrically, they are, however, very unlikely, given that the companion would spend the vast majority of the time at much larger separations from the primary star than where it was discovered. Indeed if we use the orbital period of roughly 26000\,yr of the best fitting LSMC orbit we can estimate that the probability to find the companion within 30 years of the periastron passage is only on the order of 0.1\,\%. However, the orbits that pass the periastron within the next few decades can have lower eccentricities and semi-major axes. For an eccentricity around $\sim$0.5 there is a strong peak for the time of the periastron passage in the year 2040. It will thus be increasingly interesting to continue an astrometric monitoring of this system, since significant acceleration (i.e. curvature of the orbit) would be expected especially for cases with non-extreme eccentricities. 

Since it is of specific interest if the system might indeed exhibit a high eccentricity (i.e. for plausibility of scattering scenarios during its early formation), we examined how reliable the eccentricities of our recovered orbit solutions are. \citet{pearce2014} studied the possibility that an unknown inner (sub)stellar companion could introduce a false-positive eccentricity signal in the relative astrometry between the primary star and the known directly imaged companion. This is due to the astrometric displacement of the primary star as it orbits around the common center of mass with the hypothetical inner companion. We used their formalism to calculate the mass and angular distance that would be required for such an inner companion to make the orbit solution for GJ\,758\,B appear eccentric when the real orbit is in fact circular. We did this for all the orbit solutions that fit the astrometric measurements, and the results are shown in Fig.~\ref{fig:pearce-gj758}. We find that such a hypothetical inner companion would need masses between 0.02~\MSun and 0.14~\MSun with an orbit separation of 0.42\arcsec (6.6~AU; dependent on the mass of the system and the epoch difference of the astrometric observations). For the large majority of our solutions, we can reject such a companion because it would have been discovered in our deep IRDIS images (see Sect.~\ref{sec:sensitivity_additional_companions} for detection limits estimations). However, due to the old age of the system, we would have not been able to recover inner companions with masses below $\sim$0.05~\MSun at the required angular separation. 

\begin{figure}
  \centering
  \includegraphics[width=0.5\textwidth]{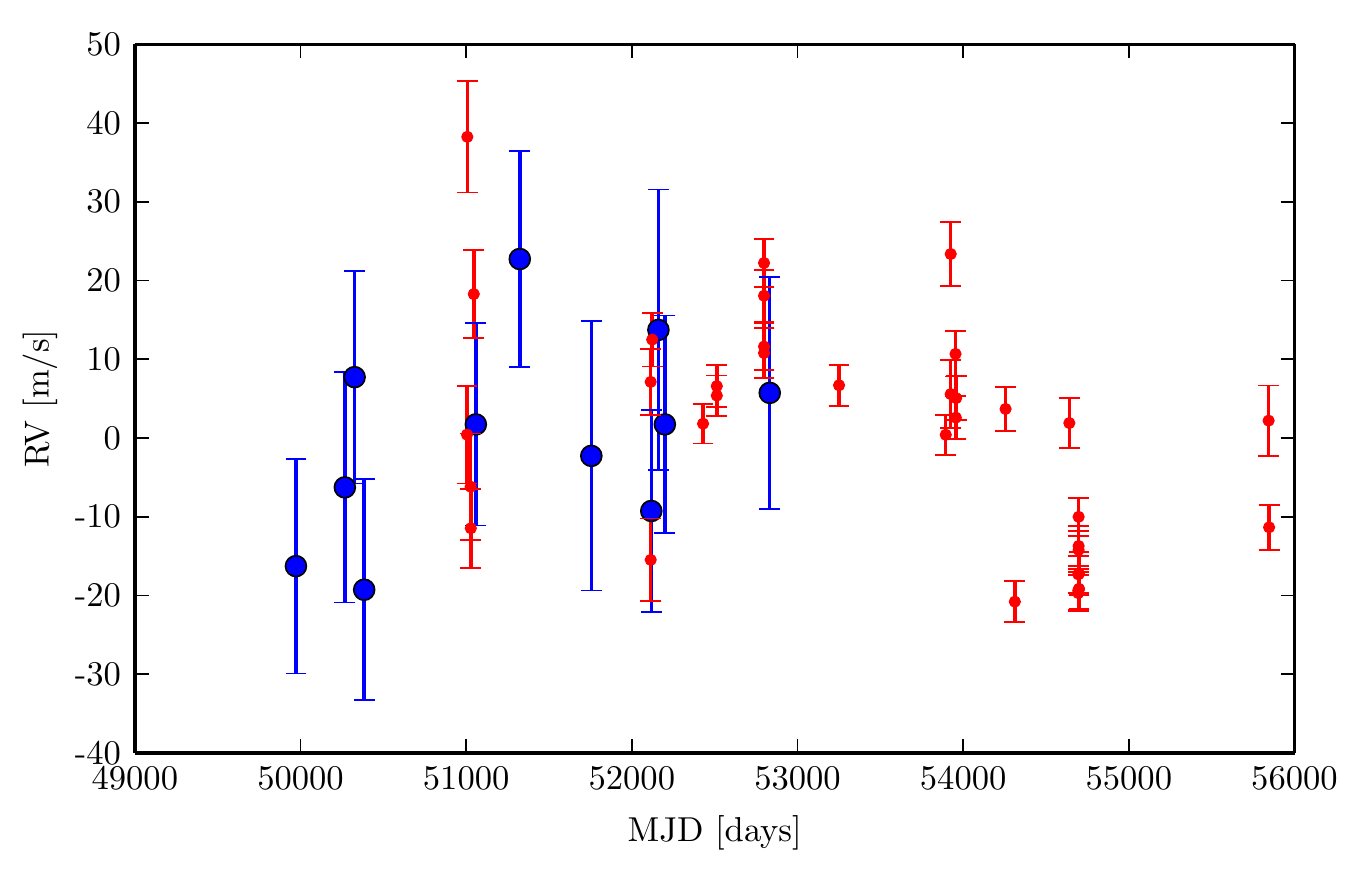}
  \caption{Radial velocity measurements of GJ\,758\,A, retrieved from the ELODIE (large blue dots) and Lick Planet Search (small red dots) archives. The measurements cover a total of 16~years which is nearly the full period of the inner companion speculated from the astrometric signal.}
  \label{fig:rvplot}
\end{figure}

To exclude the remaining possible solutions, we retrieved archival radial velocity data of GJ\,758\,A obtained with the ELODIE high-precision fiber-fed echelle spectrograph \citep{baranne1979} covering a time baseline of 7.8~years, as well as archival data from the Lick Planet Search program covering 13.2~years \citep{fischer2014}. This combined data, showed in Fig.~\ref{fig:rvplot}, covers a total of 16~years. It can be used to reject any hypothetical companion on a 17~years period more massive than 0.02~\MSun having any inclination higher than 5 degrees. Given the spherical symmetry of the system, this translates into a rejection of 98\% of the orbital solutions for a hypothetical 0.02~\MSun inner companion and an increasing rejection rate for higher masses. It is thus extremely unlikely that the observed eccentricity is due to an inner companion causing an astrometric signal.

\subsection{Markov-Chain Monte-Carlo orbital fitting}
\label{sec:mcmc_orbital_fitting}

\begin{figure*}
  \centering
  \includegraphics[width=0.8\textwidth]{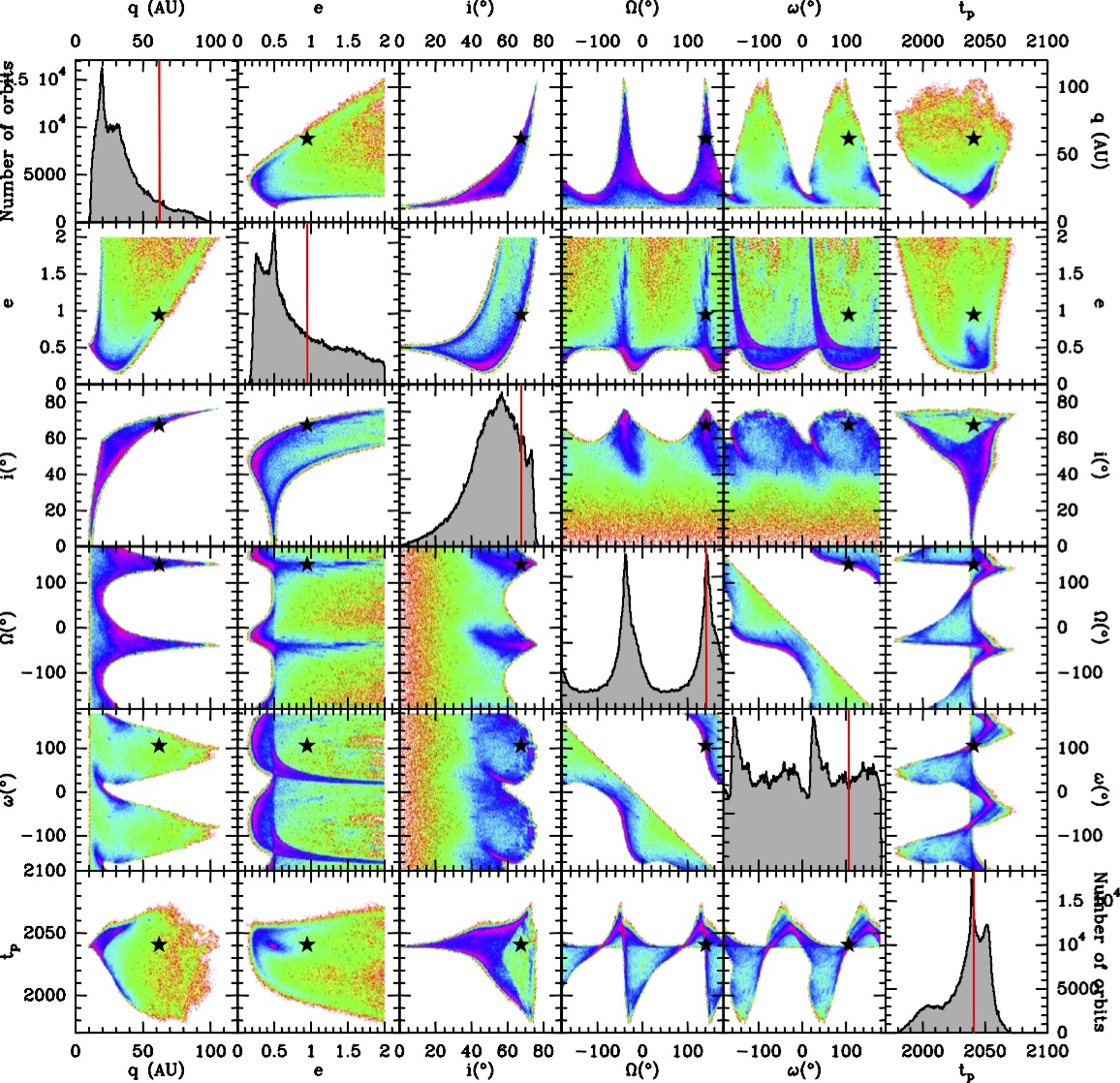}
  \caption[]{Resulting MCMC posterior distribution of the six orbital elements ($q$, $e$, $i$, $\Omega$, $\omega$, $t_p$) of GJ\,758\,B's orbit using the universal variable code. The diagonal diagrams show mono-dimensional probability distributions of the individual elements. The off-diagonal plots show bidimensional probability maps for the various couples of parameters. This illustrates the correlation between orbital elements. The logarithmic color scale in these plots is linked to the relative local density of orbital solutions. It is indicated on the side of Fig.~\ref{aeai_gj758}. In the diagonal histograms, the red bar indicates the location of the best $\chi^2$ solution obtained via standard least-square fitting. The location of this solution is marked with black stars in the off-diagonal plots. This solution is also plotted in Fig~\ref{fig:orbits}.}
  \label{mcmc_gj758}
\end{figure*}

\begin{table*}
  \centering
  \caption{Orbital characteristics of the best $\chi^2$ solution recovered by simple least-squares fitting (first column), as well as LSMC fitting (second column) for GJ\,758\,B and statistical properties of the posterior distribution. In addition, we give an example for a probable orbit (orbital elements close to MCMC peak values) that was recovered by MCMC (third column). Note that $\Omega$ and $\omega$ are defined within a $\pm180\degr$ degeneracy, so that giving confidence intervals is meaningless. Peak and confidence intervals for $a$ and $P$ in the MCMC analysis are defined for bound orbits.}
  \begin{tabular}{lcccccc}
  \hline
  \hline
  Parameter       & Best simple $\chi^2$  & Best LSMC & Probable MCMC & MCMC peak       & 67\%\ confidence & 95\%\ confidence \\
                  & solution             & solution   & solution      & value           & interval         & interval         \\
  \hline
  $a$ (AU)        & 879.29               & 878.62     & 46.05         & 33.6            & [19.7--83.7]     & [19.7--348.7] \\
  $a$ (arcsec)    & 55.79                & 55.75      & 2.922         & 2.132           & [1.250--5.311]   & [1.250--22.126] \\
  $q$ (AU)        & 61.55                & 63.26      & 21.87         & 19.84           & [9.99--38.11]    & [9.99--74.75]    \\
  $e$             & 0.93                 & 0.928      & 0.525         & 0.505           & [0.205--0.993]   & [0.133--1.78]    \\
  $i(\degr)$      & 67.37                & 68.08      & 46.05         & 56.63           & [43.6--70.3]     & [25.0--77.0]     \\
  $\Omega(\degr)$ & $-38.81$             & $-38.78$   & $-49.93$      & $-37.15\pm180$  & $\ldots$         & $\ldots$         \\
  $\omega(\degr)$ & $-73.67$             & $-73.98$   & 4.90          & $-155.47\pm180$ & $\ldots$         & $\ldots$         \\
  $t_p$ (yr AD)   & 2040.74              & 2040.98    & 2051.89       & 2039.3          & [2015.4--2051.4] & [1993.8--2059.7] \\
  $P$ (yr)        & 26073.26             & 26046.36   & 312.48        & 128.3           & [57.6--502.0]    & [57.6--4287.0] \\
  $\chi^2_{red}$  & 0.419                 & 0.417      & 1.434         & $\ldots$        & $\ldots$         & $\ldots$         \\
  \hline
 \end{tabular}
 \label{gj758_stats}
\end{table*}

The use of the Markov-Chain Monte-Carlo (MCMC) technique to fit orbits of companions, either detected by radial velocity or direct imaging, has become very popular in recent years. Concerning imaged planetary or sub-stellar companions, it was for instance successfully applied to $\beta$\,Pictoris\,b \citep{chau12,nielsen2014,macintosh2014}, Fomalhaut\,b \citep{kal13,beu14}, and to the 4 planet system of HR\,8799 \citep{pue14}.

MCMC is particularly well suited for imaged companions for which the observational follow-up usually covers only a small part of the whole orbit (due to large orbital periods). To fit GJ\,758\,B's orbit, we first used the code already used to fit $\beta$\,Pictoris\,b's \citep{chau12} and Fomalhaut\,b's \citep{beu14} orbits. But given the number of solutions at very large eccentricities that were hard to reach (the best fit solution has eccentricity $\sim$0.93; see Table~\ref{gj758_stats}) we moved to the use of another code that we developed recently, based on the use of universal Keplerian variables, with Metropolis-Hastings algorithm and using Gibbs sampling as convergence test. Universal variable formulation \citep{dan83,bur83,dan87} is an elegant way to provide a unique and continuous description of the Keplerian motion valid for any kind of orbit, either bound or unbound. The details of this code will be presented in Beust et al. (2015, in prep.). This code can handle both bound and unbound orbits, and is therefore not limited to elliptic orbits. It is well suited for very eccentric orbits. 

Finding very eccentric orbital solutions should indeed not be surprising. As was shown by \citet{pea15}, whenever astrometric orbits are followed over small orbital arcs, virtually arbitrarily eccentric solutions can be found depending on the unknown values of the $z$-coordinate and velocity along the line of sght. This situation nearly applies here. At least from a mathemetical point of view, unbound solutions should be valid as well. This motivated us to use the universal variable code.

10 chains were run in parallel until the Gelman-Rubin parameters $\hat{R}$ and $\hat{T}$ \citep{ford06} reach repeatedly convergence criteria for all parameters, i.e., $\hat{R}<1.01$ and $\hat{T}>1000$. This occurred after $5.2\times 10^8$ steps. At this point, a sample of $10^6$ orbital solutions is taken from the chains as representative for the posterior distribution of orbits. The orbital parameters considered are the periastron $q$, the eccentricity $e$, the inclination $i$ with respect to the sky plane, the longitude of ascending node $\Omega$ (counted from north), the argument of periastron $\omega$, and the time for periastron passage $t_p$. Note that we consider here the periastron instead of the semi-major axis, as the periastron assumes a continuous distribution from elliptical to hyperbolic orbits. The priors on those elements are assumed uniform for $\Omega$, $\omega$, $e$, and $t_p$, logarithmic for $q$ and $\propto\sin i$ for $i$. Combined with uniform prior for $\Omega$, the latter choice ensures a uniform probability distribution over the sphere for the direction of the orbital angular momentum vector. It must be stressed here that this choice of prior, especially concerning the eccentricity, is not dictated by physical likelyhood considerations, but rather by mathematical constraints on the sole basis on the available astrometric data. While a linear eccentricity prior between 0 and 1 can be realistic, obviously unbound orbits appear unprobable, given the age of the star. The probability of witnessing an ejection or a flyby right now is indeed very low. But as very eccentric solutions appeared to be compatible with the astrometric data, we wanted to allow the MCMC code to explore the unbound regime to estimate the actual constraints on the data and to avoid the introduction of artificial cut-offs.

The resulting posterior distribution is shown in Fig.~\ref{mcmc_gj758}, where probability histograms for individual elements are displayed as well as density maps for all possible pairs of parameters. The red bars that appear on the histogram plots, as well as the black stars in the bidimensional maps, correspond to the best $\chi^2$ solution that was derived using a least-square Levenberg-Marquardt fitting scheme before launching MCMC. This solution has a reduced $\chi^2=0.419$, but more than 80\%\ of the solutions in our posterior sample have reduced $\chi^2<1.5$. Peak values, confidence intervals, as well as details about the best $\chi^2$ solution, are given in Table~\ref{gj758_stats}. It can be seen in Fig.~\ref{mcmc_gj758} that the eccentricity distributions extend beyond $e=1$, so that we have both bound and unbound solutions in our sample. The upper limit at $e=2$ in the eccentricity distribution is not physical. This threshold was fixed at the beginning of the simulation to save computing time.

The plots involving $\Omega$ and $\omega$ appear twofold with similar patterns saparated by $\pm180\degr$. This is a direct consequence of the degeneracy of the projected astrometric motion \citep{beu14}. To each solution with ($\Omega$, $\omega$) values corresponds a twin solution with the same other orbital elements but with ($\Omega+\pi,\omega+\pi$). Both generate the same projected orbital motion.

Our first comment on the result is that the orbit is clearly eccentric. However, despite the presence of unbound solutions in our sample, and although the best $\chi^2$ solution appears very close to $e=1$, most solutions have moderate eccentricities $\la0.7$, with a peak around $e=0.5$. 68\% of orbits in our sample are bound. This is enough to stress that GJ\,758\,B is very probably a bound companion to GJ\,758, as an unbound orbits would mean an ongoing flyby or a very recent ejection. As both configurations can be regarded as improbable (though not impossible), finding more than 2/3 of bound solutions in our sample is a very strong indication for a bound orbit. 68\%\ can be regarded as the minimum probability to have a bound orbit, but the actual probability is in fact much higher.

As noted above, this value is very probably far below the actual probability, as an unbound orbit would mean an ongoing flyby or a very recent ejection. This is a very improbable configuration given the age of the star. 68\%\ can be regarded as the minimum probability to have a bound orbit without any physical consideration about the likelyhood of unbound configurations. It is thus sufficiently high to allow us to stress that GJ\,758\,B is an actually bound companion to GJ\,758\,A. Based on the ratio between the timescale of an ejection event and the age of the star, the actual probability to be observing one today should not exceed $\sim$$10^{-6}$.

The periastron lies in the range 10--40~AU for about 70\% of solutions, so that this must be regarded as the most probable range, with a clear probability peak at $q=20$~AU. Solutions with higher $q$ values mostly correspond indeed to unbound solutions, and must therefore be considered as less probable. 

All orbital solutions have inclination $i$ well below $90\degr$, compatible with a prograde motion as seen from the Earth. The inclination peaks around $~60\degr$, while the longitude of ascending node $\Omega$ exhibits a clear peak around $\sim -40\degr$. This shows that the orbital plane of GJ\,758\,B is rather well constrained. Conversely, the argument of periastron $\omega$ is very badly constrained. This is a direct consequence of the eccentricity dispersion of the solutions, as can be seen from the $(\omega,e)$ density maps (Fig.~\ref{mcmc_gj758}). The periastron passage is however better constrained, with a peak in 2039 for the next occurrence.

\subsection{Conclusion on astrometry}
\label{sec:conclusion_astrometry}

\begin{figure*}
\centering
\begin{minipage}{.30\textwidth}
  \centering
  \includegraphics[width=0.95\textwidth]{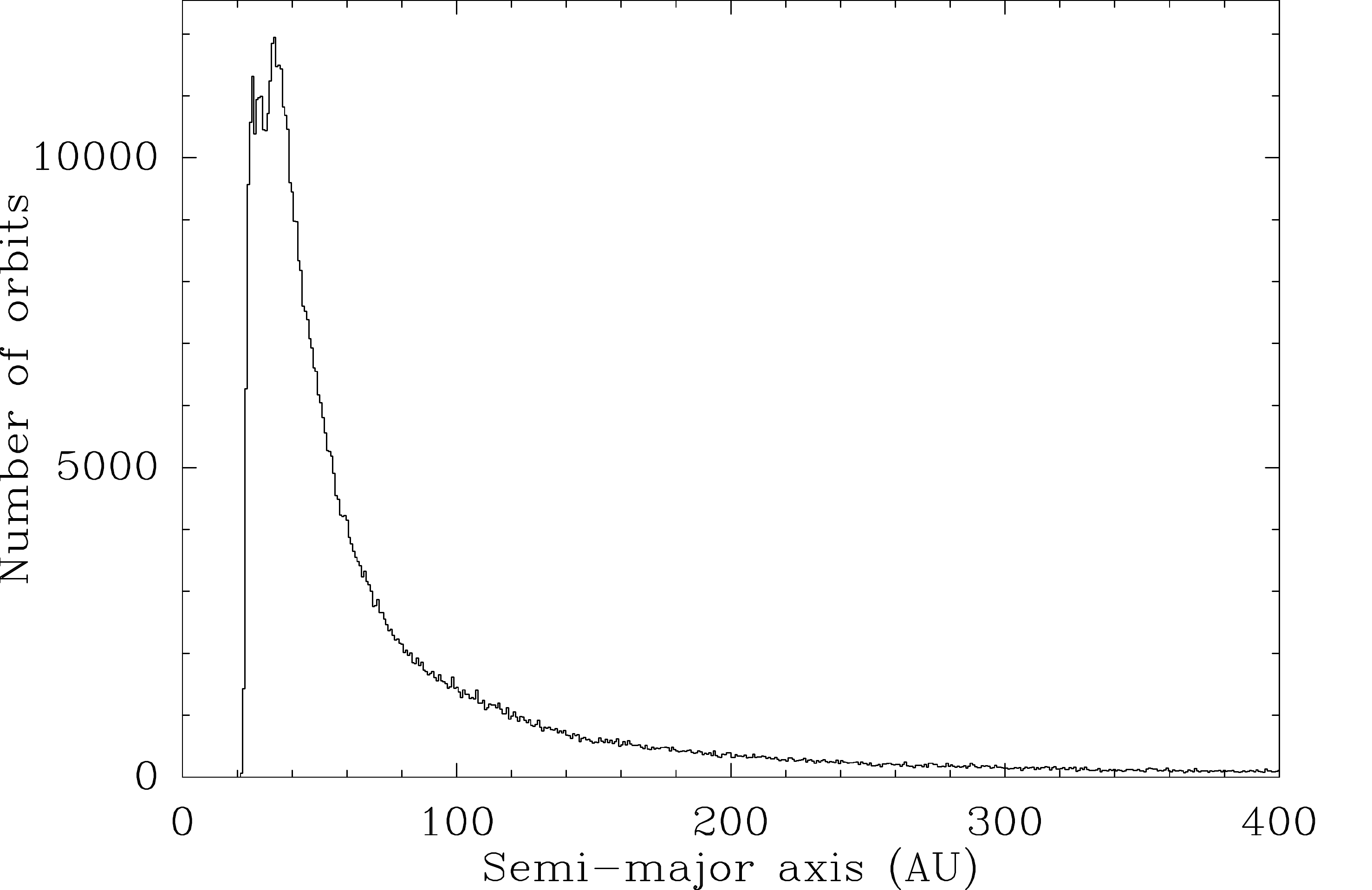}
  \includegraphics[width=0.95\textwidth]{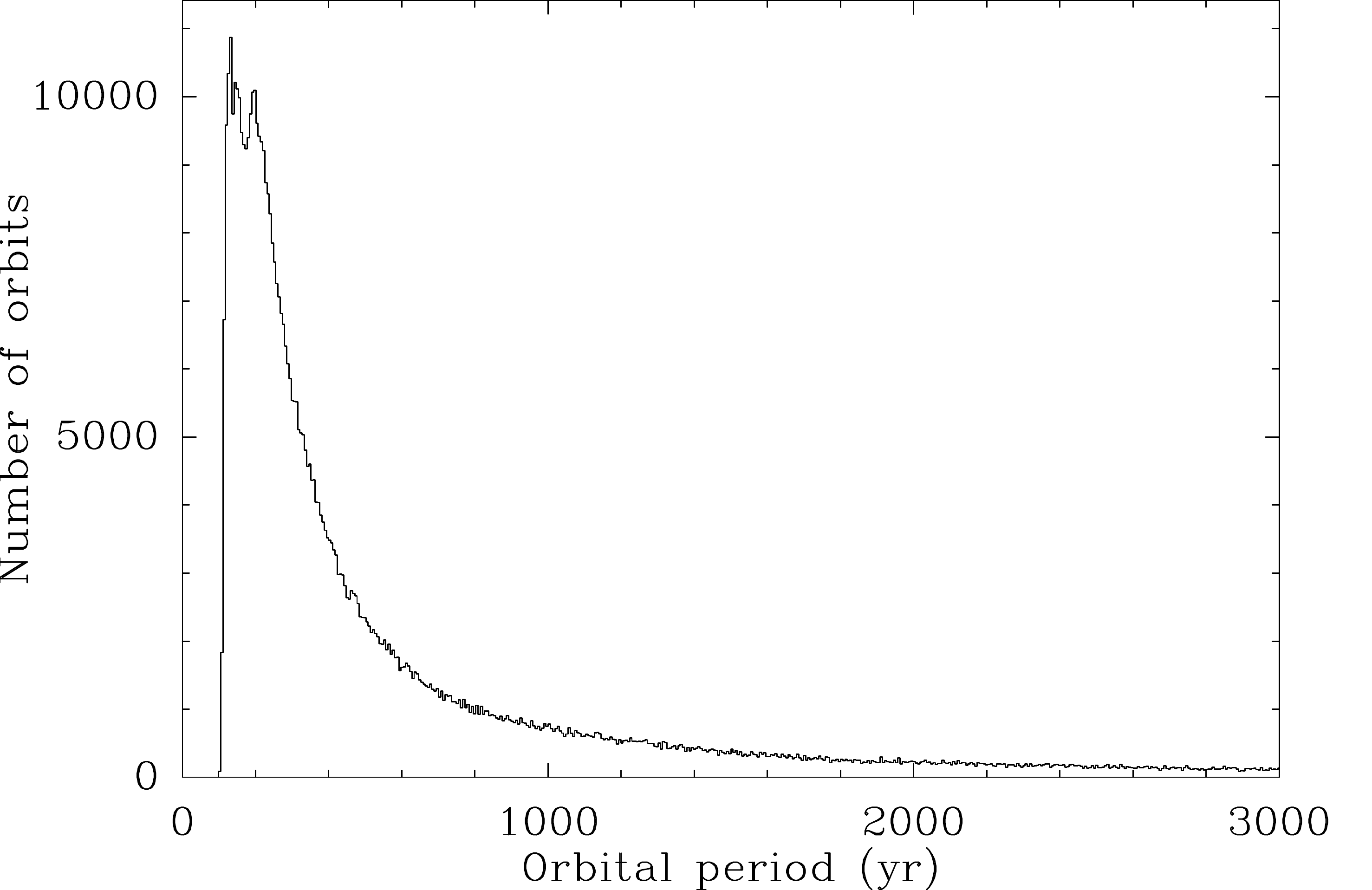}
\end{minipage}
\begin{minipage}{.30\textwidth}
  \centering
  \includegraphics[width=0.95\textwidth]{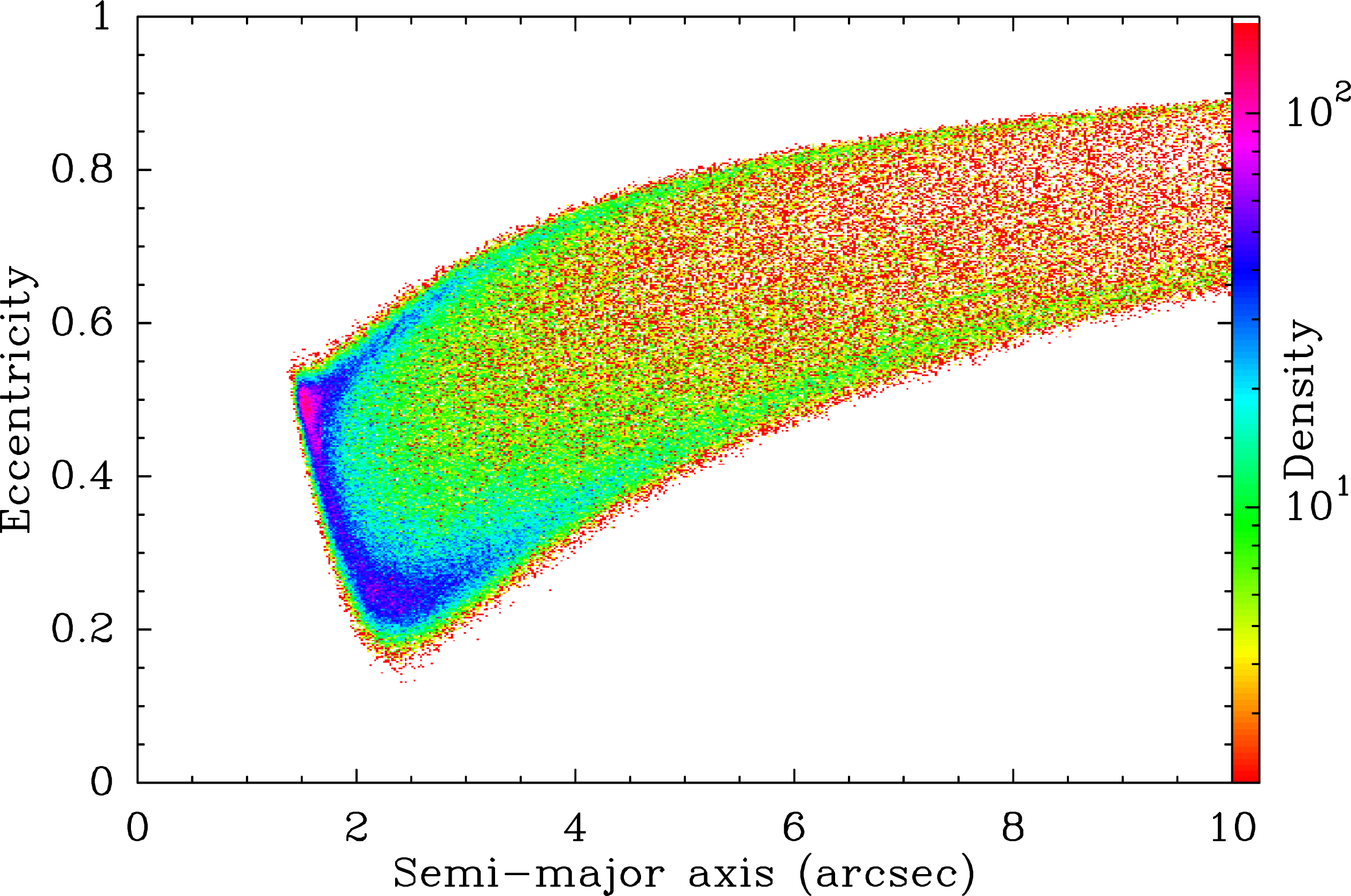}
  \includegraphics[width=0.95\textwidth]{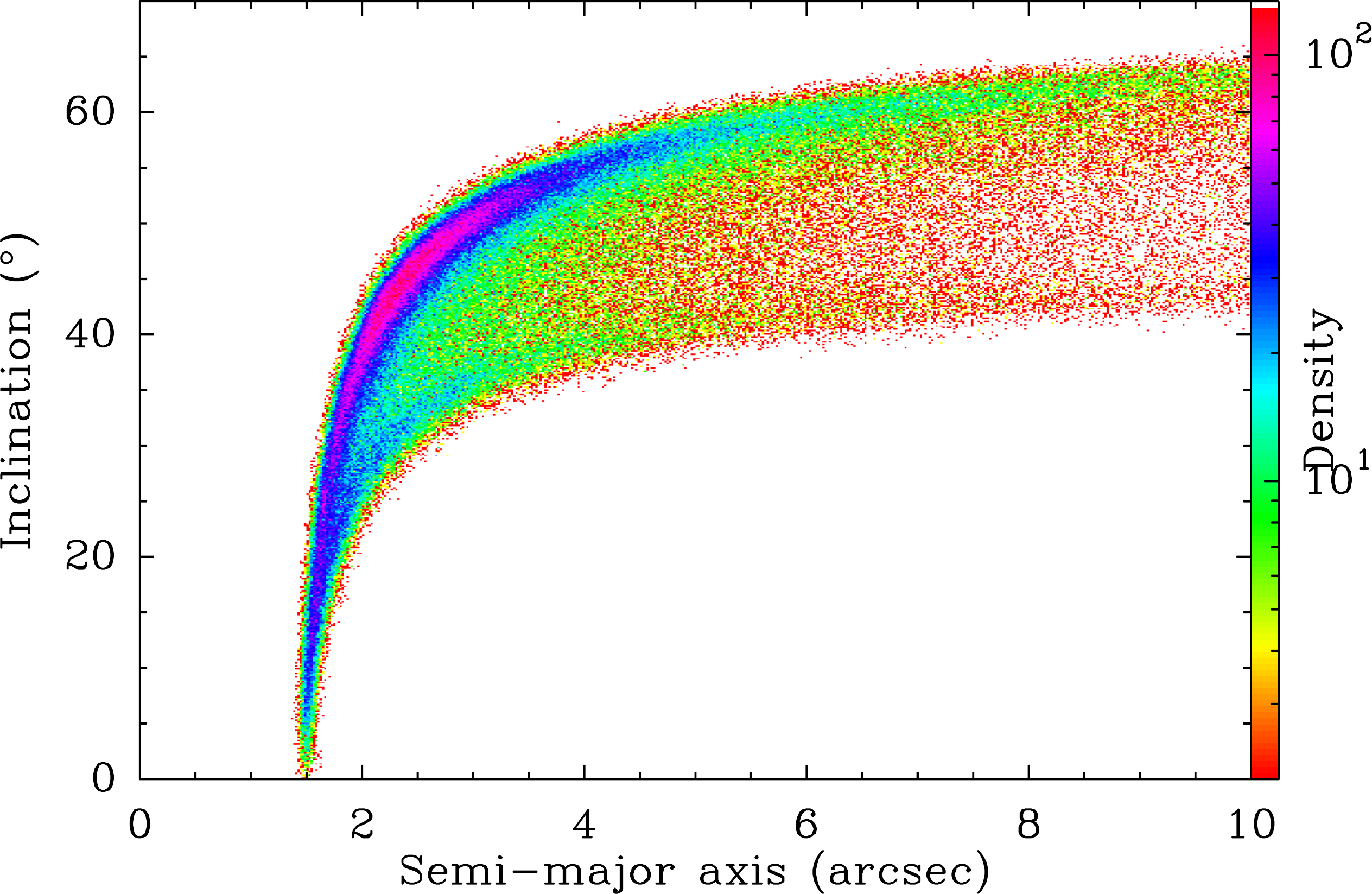}
\end{minipage}
\begin{minipage}{.30\textwidth}
  \centering
  \includegraphics[width=1.04\textwidth]{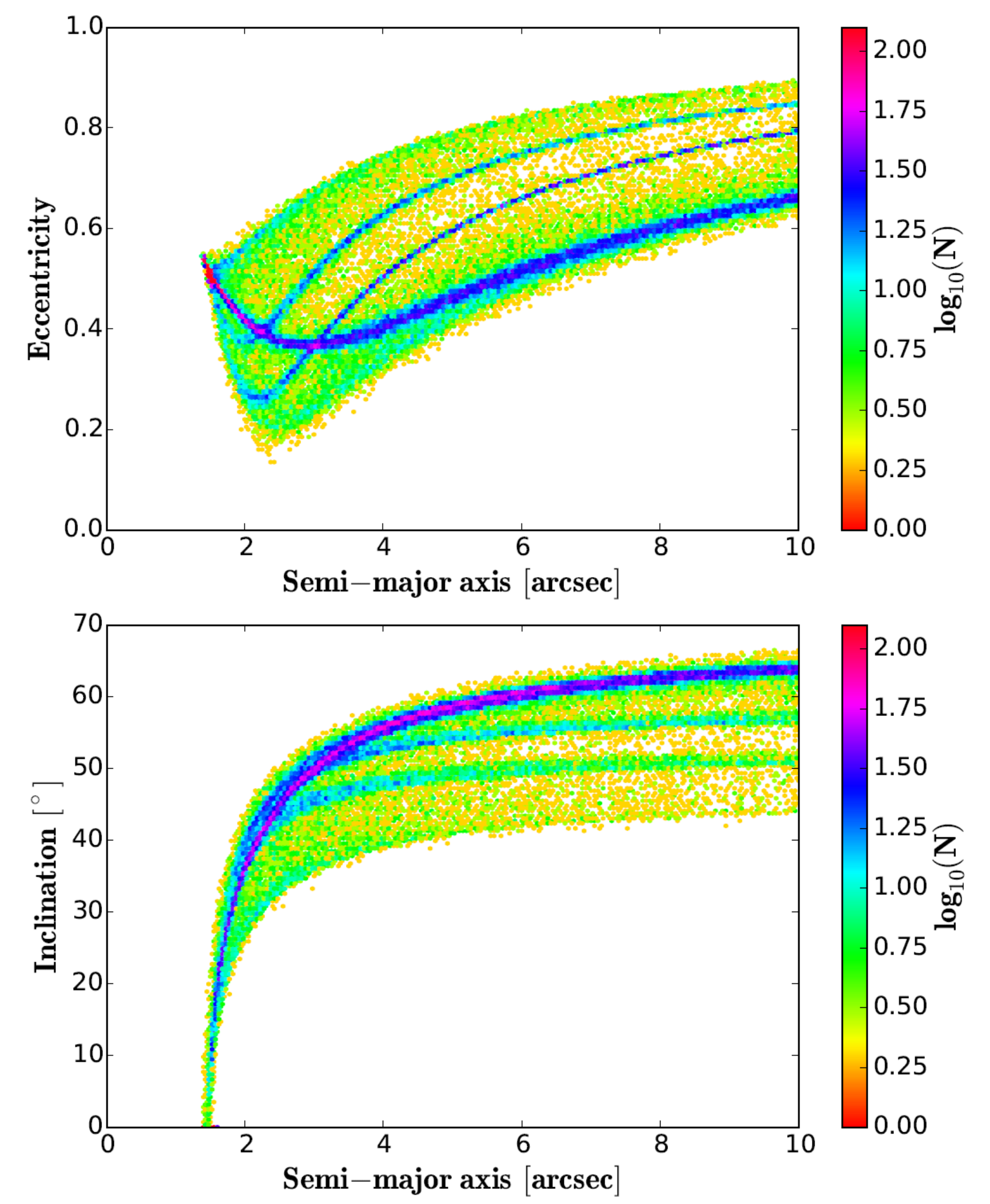}
\end{minipage}
\caption{\textit{Left and Middle:} Additional plots to Fig.~\ref{mcmc_gj758} restricted to bound orbits only recovered by our MCMC fit, showing i/ posterior distributions of semi-major axis and orbital period (\textit{left}), ii/ bidimensional density maps involving the semi-major axis $a$ versus the eccentricity $e$ and the inclination $i$. The color scale appearing on the right side of the plots also applies to all similar plots of Fig.~\ref{mcmc_gj758}. \textit{Right:} The same bidimensional density maps as shown in Fig.~\ref{aeai_gj758} showing orbit solutions recovered by our LSMC fit with semi-major axes smaller than 10\arcsec.}
\label{aeai_gj758}
\end{figure*}

To better compare LSMC and MCMC results we created matching bidimensional density maps restricted to bound orbits, considering the semi-major axis $a$ and the eccentricity $e$ as well as the inclination $i$. For the MCMC results we restricted ourselfs to bound solutions only. To match the MCMC results closely we cut off the LSMC results at semi-major axes smaller than 10\arcsec. The results are shown in Fig.~\ref{aeai_gj758}. We first show posterior MCMC distribution for the semi-major axis and the orbital period (left plots). We see that although both histograms exhibit tails towards large values (and thus approaching unbound orbits), clear peaks appears around $\sim 40\,$AU and $\sim 200\,$yr. Those values must be regarded as the most probable ones. Then we come to comparing the 2D maps generated by MCMC and LSMC (middle and right plots). This immediate comparison shows that both approaches agree very well to derive the same well-fitting range of bound orbital solutions. The differences in the density of solutions are likely caused by the difference of prior distributions used as input for both methods. While for LSMC uniform distributions in $a$ and $i$ were considered we used distributions that were uniform in $log\,q$ and $\propto\sin i$ for $i$. This was done because the aim of the methods is somewhat different. With LSMC we aim to find the full possible range of geometrically well fitting orbits, as well as the best fitting orbit in terms of $\chi^2_{red}$. With MCMC, on the other hand, the goal is to find the correct posterior probability distribution given our prior knowledge of the system. This knowledge includes that shorter period orbits are more likely given were we find the companion in our observation epochs as well as the statistical likelyhood of orbit inclinations.

We want to stress that the general results obtained by both methods agree very well. In particular with LSMC we find an upper limit of the inclination of bound orbits of 70.8~deg, while the upper limit of the 1\,$\sigma$ confidence interval recovered by LSMC is 70.3~deg. Both methods also find strong peaks in the time of the periastron passage between the years 2039 and 2040. The MCMC fit furthermore recovered 68~\% of bound orbit solutions. Given this high likelyhood of a bound orbit and the large chance for a periastron passage within the next few decades, the GJ\,758 system remains an interesting target for continued orbital monitoring. It is likely that orbit curvature will be discovered in this timeframe, allowing for a much better constrained determination of the companions orbit.

\section{Sensitivity to additional companions}
\label{sec:sensitivity_additional_companions}

\begin{figure*}
  \centering
  \includegraphics[width=1.0\textwidth]{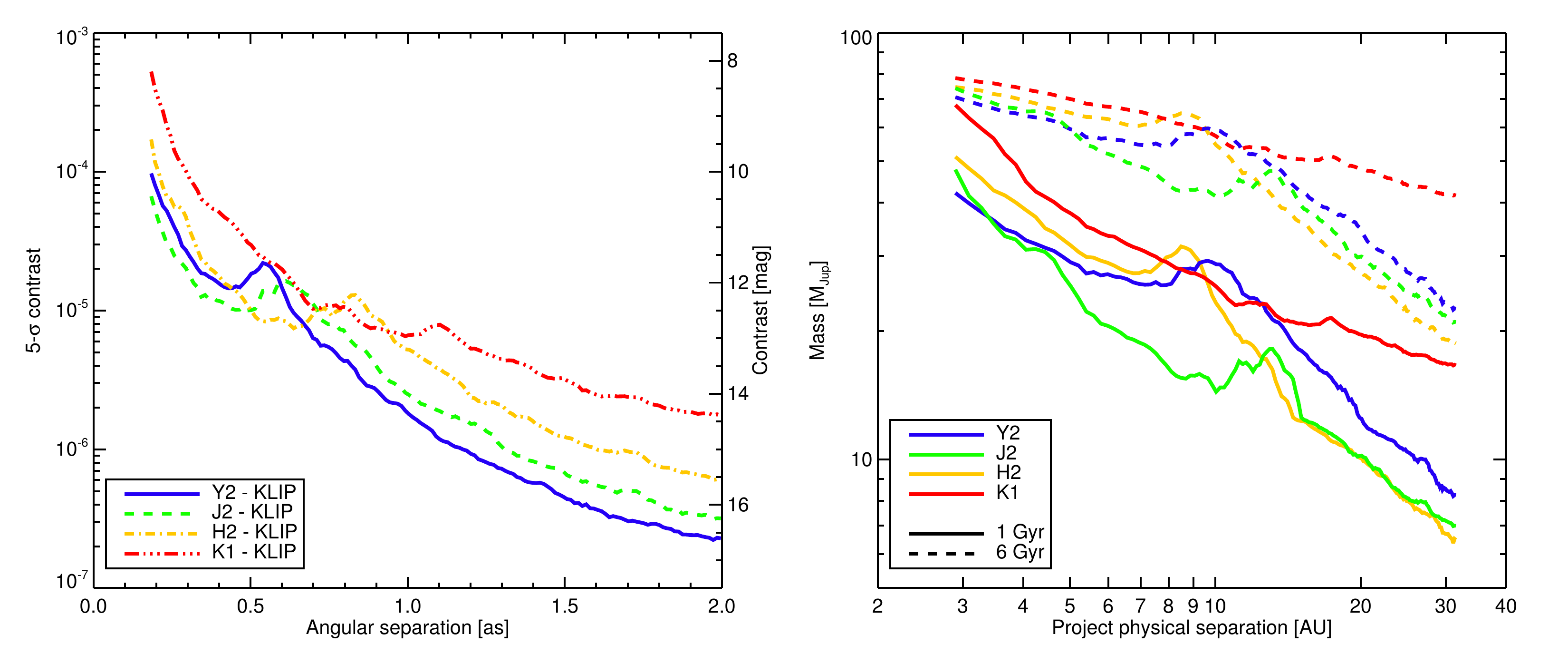}
  \caption{\emph{Left:} 5-$\sigma$ detection limits measured in the Y2, J2, H2 and K1 filters using a KLIP analysis, with a number of subtracted modes that varies to maximize the algorithm throughput. \emph{Right:} Conversion of these detection limits into physical units using the known distance of the GJ\,758 system (15.76~pc) and the AMES-COND evolutinary tracks \citep{baraffe2003} calculated in the IRDIS filters. Two sets of curves assuming the two extremes of the system age range (1 and 6~Gyr, see Sect.~\ref{sec:stellar_parameters}) are displayed. The limits for the nominal age of 3~Gyr lie in-between, slightly closer to the the 6~Gyr limit than the 1~Gyr limit.}
  \label{fig:detlim}
\end{figure*}

To conclude our analysis, it is interesting to look at our sensitivity to additional massive companions in the GJ\,758 system. We calculated detection limits in the different DBI filters following an ADI analysis with KLIP. The limits were estimated by measuring the standard deviation of the residuals in annuli of width 1~$\lambda/D$ at increasing angular separation, divided by the maximum of the off-axis PSF of the star in the same filter. To properly take into account the effect of self-subtraction induced on the detection limits, the algorithm throughput was estimated by injecting fake companions regularly spaced from 0.1\arcsec to 2.0\arcsec into the pre-processed data cubes. They were injected at a level 10 times higher than the noise residuals in the final images at the separation of each of the companions. This process was repeated 10 times with different orientations of the pattern of fake companions to average out possible variations of the throughput as a function of the position in the field. The throughput at each separation was then calculated to be the mean throughput over the 10 measurements. 

The final detection limits taking into account the throughput in the different filters are plotted in the right panel of Fig.~\ref{fig:detlim}. Due to the small amount of field rotation in all observing sequences ($\lesssim$8$^{\mathrm{o}}$), the throughput of the analysis decreases significantly towards small angular separations, resulting in a sharp deterioration of the detection limits. It is only at separations larger than 0.21\arcsec, 0.25\arcsec, 0.33\arcsec and 0.43\arcsec that the field rotates by more than $\lambda/D$ over the course of the complete sequence in $Y$-, $J$-, $H$- and $K$-band respectively. The sensitivity is nonetheless improved compared previous measurement below $\sim$0.5\arcsec, in particular for this specific target.

The right panel of Fig.~\ref{fig:detlim} shows the conversion of these detection limits into physical units of projected separation and physical mass, using the known distance of the star (15.76~pc) and the AMES-COND evolutionary models \citep{baraffe2003} calculated in the IRDIS DBI filters. The two sets of curves represent the limit for the two extremes of the system's age range, 1--6~Gyr (see Sect.~\ref{sec:stellar_parameters}). For the youngest part of the age range, our observations clearly probe the low mass brown-dwarf regime down to 4--5~AU, and even the planetary-mass regime beyond 20--30~AU. However, if we assume an older age for the system, more in line with the different age indicators, only massive brown-dwarfs could be detected. 

\section{Summary \& conclusions}
\label{sec:summary_conclusions}

Our new study of GJ\,758 offers an improved overview of this interesting system. The brown-dwarf companion is redetected and we confirm some of its already known properties using our finer spectral sampling from SPHERE/IRDIS observations. In particular, we recover a low $\Teff = 600 \pm 100$~K from comparison to 4 different sets of models, in good agreement with previous studies \citep{thalmann2009,currie2010,janson2011}. There are however some interesting peculiarities that are worth discussing.

From comparison to empirical objects, GJ\,758\,B appears as a very interesting object because we cannot find objects with known age and metallicity that matches all of its observed fluxes. We estimate a T8 spectral type for this object, but this estimation is limited by the small amount of spectra of T8--T9 dwarfs with robust constraints on their age and metallicity. One of the most likely explanations for the peculiar SED of this brown dwarf is the super-solar metallicity of the primary (${\rm[Fe/H]} = 0.18 \pm 0.05$; see Sect.~\ref{sec:abu}), and of the companion if we assume they share the same metal enrichment. This is supported by the fact that the T8pec companion to the metal-poor dwarf BD+01\,292 \citep{2012MNRAS.422.1922P} shows a $K$-band spectral deviation opposite to that of GJ\,758\,B. Unfortunately, the current lack of such companions precludes us to perform a meaningful comparison. Similarly, comparison to synthetic grids of models are constrained by the limited extension of most grids toward non-solar metallicities and low \Teff values. None of the four types of models that we tested were able to reproduce accurately the photometry of GJ\,758\,B in all filters. Especially, the J3 and H2 fluxes are very hard to reproduce and cannot be readily explained with any of the models. The BT-SETTL14 models \citep{2013MSAIS..24..128A}, with an enrichment of 0.3~dex in $\alpha$ elements with respect to solar, do provide a better fit of the J3 flux, but at the price of much smaller dilution factors corresponding to unphysical values of the companion radius ($\sim$0.7~\RJup). Even so, the high H2 flux is not reproduced. This could actually indicate that the \Teff of GJ\,758\,B might be even lower than the one of the best fit, but we are limited by the absence of metal-enriched models at very low \Teff. As a result, our analysis confirms the low \Teff of the object, corresponding to a mass of $23^{+17}_{-13}$~\MJup in the considered age range (using the \citealt{baraffe2003} evolutionary tracks), but we cannot infer any precise value for \logg and M/H. The study of this object would strongly benefit from low-resolution spectroscopy, e.g. with IRDIS long-slit spectroscopy mode \citep{vigan2008}.

The new astrometry confirms the picture of a very eccentric companion. Similarly to what was reported by \citet{janson2011}, the curvature of the orbit is still not detectable yet. The $\sim$0.28\arcsec motion of the companion along its orbit appears roughly as a straight line from the previous measurements from 2010. Our LSMC and MCMC simulations favour an eccentric -- but bound -- orbit, with high-likelihood of $e \simeq 0.5$. In particular, no orbital solution shows an eccentricity lower than 0.14, which is consistent with previously reported results and between our two approaches ([0.133--1.78] 95\% confidence interval from MCMC). In addition, we have ruled out the possibility that the observed eccentricity is just caused by a massive closer-in companion that would create a false positive eccentricity by astrometric displacement of the primary. Indeed, although not extremely accurate, our RV data reject companion more massive then 0.02~\MSun on periods shorter than 17~years and inclination larger than 5~deg. Finally, our new IRDIS observations reject the possibility of an additional companion more massive than $\sim$30-40~\MJup (for ages of 1-3~Gyr) above 4-5~AU.

In light of our constraints on the orbit and mass of GJ\,758\,B, it is interesting to look into the formation of this object. While our study does not bring enough new material in favor or against planet-like formation scenarios (core accretion vs. gravitational instability, migration, etc), we can instead focus on stellar-like formation scenarios. Several past studies have argued for a universal companion mass function (CMF) for stellar and sub-stellar companions. In particular, in their in-depth anaysis focused on brown dwarfs around solar-type stars, \citet{metchev2009} find tentative evidence for such a universal CMF and predict a peak in semi-major axes for brown dwarfs at $\sim$30~AU. From that point of view, it makes sense to compare the properties of the GJ\,758 system to the properties of solar-type multiple systems. \citep{raghavan2010} have published the most complete study on this topic to date. While we cannot compare quantitatively our results on a single object with their global multiplicity analysis, it is interesting to note a few qualitative facts. Firstly, with a most likely period of 312.48~yr (from MCMC, see Table~\ref{gj758_stats}), GJ\,758\,B falls exactly at the peak of their period distribution (293.57~yr; their Fig. 13). Secondly, with a most likely eccentricity of 0.525, GJ\,758\,B falls in the bulk of their eccentricity--period distributions. And thirdly, with a ratio $q = M_{2} / M_{1} = 0.023 \pm 0.013$, the GJ\,758 systems falls at the very edge of the study of \citep{raghavan2010}, but in their $q$--period plot, the system is not completely isolated and could be part of the tail of the distribution. While these facts do not prove that the system was formed in a stellar way, they certainly support the possibility qualitatively.

In conclusion, GJ\,758\,B remains a very interesting object that warrants deeper observations to look for additional companions in the system, spectroscopic observations to better constrain its physical properties, and astrometric monitoring to get tighter constraints on its eccentric orbit.

\begin{acknowledgements}
AV, MB, GC, GS, JLB and DM acknowledge support in France from the French National Research Agency (ANR) through project grant ANR10-BLANC0504-01, the CNRS-D2P PICS grant, and the Programmes Nationaux de Planetologie et de Physique Stellaire (PNP \& PNPS). JLBa's Ph.D is funded by the LabEx Exploration Spatiale des Environnements Plan\'etaires (ESEP) \#2011-LABX-030. VD is partially supported by the Australian Research Council. VD, SD, ALM, RG, and DM acknowledge support from the ``Progetti Premiali'' funding scheme of the Italian Ministry of Education, University, and Research. EB and JH are supported by the Swiss National Science Foundation (SNSF). AZ acknowledges support from the Millennium Science Initiative (Chilean Ministry of Economy), through grant ``Nucleus RC130007''. The authors warmly thank A. Bellini and J. Anderson for kindly providing the catalog positions of the stars in the \object{47\,Tuc} field before their publication. We are very grateful to D. Kirkpatrick, A. Burgasser, D. Pinfield, B. Burningham, and G. Mace for sending us the spectra of benchmark T--Y objects. We thank C. Morley, D. Saumon and F. Allard for the free online access to their atmospheric model grids, and F. Bouchy for his help with the ELODIE data. This research has benefitted from the SpeX Prism Spectral Libraries, maintained by Adam Burgasser at \url{http://pono.ucsd.edu/~adam/browndwarfs/spexprism}. This research has made use of the SIMBAD database, operated at CDS, Strasbourg, France. \\

SPHERE is an instrument designed and built by a consortium consisting of IPAG (Grenoble, France), MPIA (Heidelberg, Germany), LAM (Marseille, France), LESIA (Paris, France), Laboratoire Lagrange (Nice, France), INAF -- Osservatorio di Padova (Italy), Observatoire de Gen\`eve (Switzerland), ETH Zurich (Switzerland), NOVA (Netherlands), ONERA (France) and ASTRON (Netherlands), in collaboration with ESO. SPHERE was funded by ESO, with additional contributions from CNRS (France), MPIA (Germany), INAF (Italy), FINES (Switzerland) and NOVA (Netherlands). SPHERE also received funding from the European Commission Sixth and Seventh Framework Programmes as part of the Optical Infrared Coordination Network for Astronomy (OPTICON) under grant number RII3-Ct-2004-001566 for FP6 (2004-2008), grant number 226604 for FP7 (2009-2012) and grant number 312430 for FP7 (2013-2016).
\end{acknowledgements}

\appendix

\section{Description of the atmospheric models}
\label{sec:description_atmospheric_models}

The specificity of the models in the range of \Teff suited for GJ\,758\,B are not all described in the literature yet. So it is important to make a description of the most relevant hypothesis in the models and differences between the models in this paper. The parameter space of the models is summarised in Table~\ref{Tab:atmomodchar}.

The BT-Settl model couples a cloud model to a 1D radiative transfer code \texttt{PHOENIX} \citep{1994ApJ...426L..39A, 1997ApJ...483..390H}. The model considers the formation of a cloud deck composed of up to 55 grain species. The grain size and density, the abundances of chemicals in the gas phase, including the effect of element depletion induced by the grain formation, is computed layer per layer through the photosphere following a comparison of the timescales for nucleation, condensation, gravitational settling or sedimentation, and mixing. Once rained out below the photosphere, the grain opacities are not accounted for into the radiative transfer. Nevertheless, these grains can still interact chemically with the gas phase. These models can predict the flux at the surface of a given object only defined by \logg, \Teff, and [Fe/H]. They account for the non-equilibrium chemistry of CO, CH$_{4}$, N$_{2}$, NH$_{3}$, and CO$_{2}$. The models predict the formation of  a secondary (resurgent) cloud layer into the photosphere made of Na$_2$S and MnS then of KCl, NaCl and some ZnS that lies above the rained-out primary cloud layer located below the photosphere and that originally sustains in the atmosphere of L and early-T dwarfs. Here we used the 2014 releases of the models (hereafter BT-SETTL14) which include revised  alkali cloud opacities and the latest CIA opacities \citep{2012APS..MARY26001A}. A specific grid was computed for the project to cover the Y-dwarf temperature domain (hereafter BT-SETTL14-Y) in addition to the already existing grid covering a broader interval of \Teff and considering $\alpha$-element enrichment (hereafter BT-SETTL14). 

The 1D Exo-REM models \citep{2015arXiv150404876B} propose a simplified approach of sub-stellar atmospheres. They predict the equilibrium-temperature profile and mixing-ratio profiles of the most important gases (H2-He collision-induced absorption, H$_{2}$O, CO, CH$_{4}$, NH$_{3}$, VO, TiO, Na, and K). The absorption by iron and silicate cloud particles is added above the expected condensation levels with a fixed scale height and a given optical depth at some reference wavelength. For the purpose of the GJ\,758\,B study, two grids of models -- NC and T3 -- were computed. The NC models consider photospheres with no cloud opacity. The T3 models consider a photosphere with 30~\mic  Iron (Fe) and Forsterite (Mg$_{2}$SiO$_{4}$) grains and an optical depth of reference $\mathrm{\tau_{cloud}} = 3$. The grains are located between condensation level and a 100 times lower pressure. They  have scale heights equal to the gas scale height, and optical depths of 3 and 0.45 at 1.2~\mic respectively for Fe and Mg$_{2}$SiO$_{4}$.

Similarly to the BT-Settl models, the Morley+12 models account for the possible resurgence of clouds in late-T dwarf atmospheres. These 1D models build on the cloud model of \cite{2001ApJ...556..872A}. The cloud content (and opacity) is determined by a balance between the upward transport by turbulent mixing with the sedimentation. The \cite{2001ApJ...556..872A} models do not compute the microphysics associated to the clouds, and rather let as free parameters the vertical eddy diffusion coefficient $K_{zz}$ and a sedimentation efficiency parameter $f_{sed}$. A higher $f_{sed}$ correspond to thinner (rained-out) clouds. For the case of the Morley+12 models, non-equilibrium chemistry is not included, so only models with $K_{zz}=0$ are available. The models only enable an exploration of the $f_{sed}$. As a second difference with the BT-Settl, the Morley+12 models do not account for chemical reactions between the condensed species and the gas phase. Finally, we added to this grid the models of \cite{2012ApJ...750...74S}, also based on \cite{2001ApJ...556..872A} models, to explore the case of an extreme sedimentation efficiency (cloud-free atmospheres). 

\begin{table*}[t]
\centering
\caption{\label{Tab:atmomodchar} Characteristics of the atmospheric model grids adjusted on the SED of GJ\,758\,B. [$\alpha$] stands for the $\alpha$ elements enhancement with respect to solar \citep{2011SoPh..268..255C}.}
\begin{tabular}{lllllllll}
 \hline
 Model name   & \Teff     & $\Delta\Teff$ & \logg & $\Delta\logg$ & [M/H] & $[\alpha]$ & $f_{SED}$  \\
              & (K)       & (K) & (dex)   & (dex)  & (dex)  & (dex)  \\
 \hline
 BT-SETTL14-Y & 200-420   &  20 & 3.5-4.5 & 0.5 & 0.0 & 0.0 & n/a \\
 BT-SETTL14-Y & 450-1000  &  50 & 3.5-4.5 & 0.5 & 0.0 & 0.0 & n/a \\
 BT-SETTL14   & 500-3000  &  50 & 3.5-5.5 & 0.5 & 0.0 & 0.0 & n/a \\
 BT-SETTL14   & 500-2800  & 100 & 4.0-5.5 & 0.5 & 0.0 & 0.3 & n/a  \\ 
 \hline 
 Exo-REM - NC & 400-1300  & 100 & 3.5-5.5 & 0.2 & -0.5,0,+0.5 & 0.0 & n/a \\
 Exo-REM - T3 & 400-1300  & 100 & 3.5-5.5 & 0.2 & -0.5,0,+0.5 & 0.0 & n/a \\
 \hline
 Morley+12    & 400-600   &  50 & 4.0-5.5 & 0.5 & 0.0 & 0.0 & 2,3,4,5  \\
 Morley+12    & 600-1200  & 100 & 4.0-5.5 & 0.5 & 0.0 & 0.0 & 2,3,4,5  \\
 Saumon+12    & 300-350   &  50 & 3.75,4.0-5.0 & 0.25,0.5 & 0.0 & 0.0 & $\infty$ \\
 Saumon+12    & 300-1200  &  50 & 3.0-4.75,5.0-5.5 & 0.25,0.5 & 0.0 & 0.0 & $\infty$ \\
 Saumon+12    & 1300-1500 & 100 & 4.0-5.0 & 0.5 & 0.0  & 0.0 & $\infty$ \\
 \hline
\end{tabular}
\end{table*}

\bibliographystyle{aa}
\bibliography{paper}

\end{document}